\documentclass[numerical,superscriptaddress,nofootinbib,showpacs,showkeys,longbibliography]{revtex4-1}

\usepackage{graphicx,color}
\usepackage{amsmath,amssymb,amsfonts}
\usepackage[colorlinks=true,linkcolor=blue,plainpages=false]{hyperref}

\newcommand{\be}{\begin{equation}}
\newcommand{\ee}{\end{equation}}
\newcommand{\ba}{\begin{eqnarray}}
\newcommand{\ea}{\end{eqnarray}}
\newcommand{\ban}{\begin{eqnarray*}}
\newcommand{\ean}{\end{eqnarray*}}
\newcommand{\nn}{\nonumber}

\begin{document}

\title{Influence of finite volume effect on the Polyakov Quark-Meson model}
\medskip

\author{Niseem~Magdy} 
\email{niseemm@gmail.com}
\affiliation{Department of Physics, University of Illinois at Chicago, Chicago, Illinois 60607, USA}

\begin{abstract}

In the current work, we study the influence of a finite volume on $2+1$ $SU(3)$ Polyakov Quark-Meson model (PQM) order parameters, (fluctuations) correlations of conserved charges and the quark-hadron phase boundary. Our study of the PQM model order parameters and the (fluctuations) correlations of conserved charges indicates a sizable shift of the quark-hadron phase boundary to higher values of baryon chemical potential ($\mu_{B}$) and temperature ($T$) for decreasing the system volume. The detailed study of such effect could have important implications for the extraction of the (fluctuations) correlations of conserved charges of the QCD phase diagram from heavy ion data.

\end{abstract}


\maketitle

%
\section{INTRODUCTION}\label{sec:I}
One of the major aims of the current heavy ion collisions research is to study the properties of the strongly interacting matter created in such collisions theoretically and experimentally. On the experimental level, many facilities have been designed to investigate the strongly interacting matter phase diagram such as Relativistic Heavy Ion Collider (RHIC) program ~\cite{Fachini:2006ch}, and the Large Hadron Collider (LHC) ~\cite{Monteno:2007gi}. Different studies suggest that the strongly interacting matter phase transition from hadronic phase to quark-gluon plasma (QGP) phase be a smooth crossover at low density and high-temperature~\cite{Aoki:2006we}, and first-order phase transition at high density and low temperature~\cite{Ejiri:2008xt, Pisarski:1983ms}. Both smooth crossover and first-order phase transitions expected to be connected by the critical endpoint (CEP), at which the phase transition is expected to be second order. One avenue to map out and study the QCD phase diagram is through the effective models such as the quark-meson (QM) model~\cite{Chiral:dynamics,Kovacs:2006ym,Kovacs:2007sy}, the Nambu-Jona- Lasinio (NJL) model~\cite{Nambu:1961tp}, and their Polyakov-loop extended versions~\cite{Fukushima:2003fw}.

Many studies have been devoted to investigating the QCD phase diagram, higher order moments and the thermodynamics of two~\cite{Kahara:2008yg,Wambach:2009ee} and three quark flavors~\cite{Schaefer:2008ax} QM model and even PQM model with different Polyakov–loop potentials. The thermodynamic properties (pressure, the equation of state, the speed of sound, specific heat, trace anomaly, and the bulk viscosity) have been evaluated at finite and vanishing chemical potential~\cite{Schaefer:2008ax,Mao:2009aq}. 

The effect of a finite-volume on the strongly interacting matter has been widely studied~
\cite{Fisher:1972zza,Abreu:2006pt,Palhares:2009tf,Fraga:2011hi,Bhattacharyya:2012rp,Bhattacharyya:2014uxa,Magdy:2015eda,Almasi:2016zqf}. Those studies include that finite volume has a strong effect on the transition temperature ($T_c$), the location of the critical endpoint and other thermodynamic properties. In PQM model the transition temperature ($T_c$) shifted to large values as the volume decrease \cite{Magdy:2015eda} and the location of the critical endpoint is shifted toward large $\mu$ and small $T$~\cite{Palhares:2009tf,Fraga:2011hi}. On another hand NJL \cite{Abreu:2006pt} and PNJL \cite{Bhattacharyya:2012rp} indicates that the transition temperature ($T_c$) shifted to small values as the volume decrease and the location of the critical endpoint is shifted toward large $\mu$ and small $T$ for (2+1 flavor) and toward small $\mu$ and small $T$ for (2 flavors).

In this work, we investigate the effect of the finite volumes on the PQM model order-parameters, phase-transition, and the conserved charges fluctuations and correlations. The present work is organized as follows. In section \ref{sec:II} we give a brief overview of the PQM model. The PQM model calculations of the order-parameters, thermodynamic properties, and the conserved charges fluctuations and correlations are compared with the LQCD \cite{Borsanyi:2013bia,Bazavov:2012jq}, also, the influence of finite-volume effect on the PQM model conserved-quantities, baryon, charge, strangeness and them correlations will be presented in section \ref{sec:III}. We conclude with a summary and an outlook in section \ref{sec:IIII}.

\section{The Polyakov Quark Meson (PQM) model} \label{sec:II}

The $SU(3)$ Quark Meson model with $N_f = 2+1$ flavor quarks, coupled to Polyakov loop dynamics to formulate
the Polyakov Quark Meson (PQM) model \cite{Schaefer:2008ax}. The related Lagrangian is given as;

\begin{eqnarray} \label{plsm}
\mathcal{L}=\mathcal{L}_{\rm chiral}-\mathbf{\mathcal{U}}(\phi, \phi^*, T),
\end{eqnarray}

where the chiral part of the Lagrangian, $\mathcal{L}_{\rm chiral}=\mathcal{L}_{quark}+\mathcal{L}_{meson}$, has $SU(3)_{L}\times SU(3)_{R}$ symmetry \cite{Lenaghan:2000ey,Schaefer:2008hk}. The first part provides the fermionic sector, and the second part represents the mesonic contribution, both contributions had been extensively discussed in Ref.~\citep{Mao:2009aq}.

The second term in Eq. (\ref{plsm}), $\mathbf{\mathcal{U}}(\phi, \phi^*, T)$, represents the Polyakov--loop effective potential \cite{Polyakov:1978vu}, which is expressed by using the dynamics of the thermal expectation value of a color traced Wilson loop in the temporal direction  

\begin{eqnarray}
\Phi (\vec{x})=\frac{1}{N_c} \langle \mathcal{P}(\vec{x})\rangle ,
\end{eqnarray}

Then, the Polyakov--loop potential and its conjugate read:
\begin{eqnarray}
\phi &=& (\mathrm{Tr}_c \,\mathcal{P})/N_c, \label{phais1}\\ 
\phi^* &=& (\mathrm{Tr}_c\,  \mathcal{P}^{\dag})/N_c, \label{phais2}
\end{eqnarray}
where $\mathcal{P}$ is the Polyakov loop.  This can be represented by a matrix in the color space \cite{Polyakov:1978vu} 
\begin{eqnarray}
 \mathcal{P}(\vec{x})=\mathcal{P}\mathrm{exp}\left[i\int_0^{\beta}d \tau A_4(\vec{x}, \tau)\right],\label{loop}
\end{eqnarray}
where $\beta=1/T$ is the inverse temperature and $A_4 = i A^0$ is called Polyakov gauge \cite{Polyakov:1978vu,Susskind:1979up}.

In case of no quarks, zero quark chemical potential, then $\phi = \phi^{*}$ and the Polyakov loop is recognized as an order parameter for the deconfinement phase-transition ~\cite{Ratti:2005jh}. In the present work, we used Polyakov loop effective potential $U(\phi, \phi^{*},T)$ as discussed in Refs.~\cite{Ratti:2005jh,Ghosh:2007wy} but with a new dimensionless parameter $K$ that help us get a better agreement with the LQCD.  Other Polyakov loop potentials~\cite{Haas:2013qwp,Schaefer:2009st} were also examined in various work. However, the particular selection made for this work does not affect the main conclusions of our work.
\begin{eqnarray}\label{Uloop}
\dfrac{\mathcal{U}(\phi, \phi^*, T)}{T^{4}} &=& -\dfrac{B}{2} \phi \phi^{*} - \dfrac{a_{1}}{6} ( \phi^{3} + \phi^{*^{3}}) + \dfrac{a_{2}}{4} ( \phi \phi^{*})^{2} - K~ln[1 - 6 \phi \phi^{*} + 4 ( \phi^{3} + \phi^{*^{3}}) - 3 ( \phi \phi^{*})^{2}],
\end{eqnarray}

where $B$ and $K$ are dimensionless parameters given as:

\begin{eqnarray}\label{BK}
B &=& b_{0} + b_{1}~\left( \dfrac{T_{0}}{T} \right)~ + b_{2}~\left( \dfrac{T_{0}}{T} \right)^2~ + b_{3}~\left( \dfrac{T_{0}}{T} \right)^3 ,\\
K &=& k_{1}~\left( \dfrac{T_{0}}{T} \right)~ + k_{2}~\left( \dfrac{T_{0}}{T} \right)^2~ + k_{3}~\left( \dfrac{T_{0}}{T} \right)^3 + k_{4}~\left( \dfrac{T_{0}}{T} \right)^4. 
\end{eqnarray}

 The mean field approximation is used following Refs.~\citep{Mao:2009aq,Tawfik:2014uka} to obtain the grand potential as:
\begin{eqnarray}
\Omega(T, \mu_f) &=& U(\sigma_x, \sigma_y)+\mathbf{\mathcal{U}}(\phi, \phi^*, T) + \Omega_{\bar{\psi}\psi} (T,\mu_f;\phi,\phi^{*}), 
\label{potential}
\end{eqnarray}

where $\sigma_x$ and $\sigma_y$ are the non-strange and strange chiral condensates, the first term in Eq. (\ref{potential}) is a purely mesonic potential expressed as:
%
\begin{eqnarray}
U(\sigma_x, \sigma_y) &=& \frac{m^2}{2} (\sigma^2_x+\sigma^2_y)-h_x
\sigma_x-h_y \sigma_y-\frac{c}{2\sqrt{2}} \sigma^2_x \sigma_y \nonumber \\
&+& \frac{\lambda_1}{2} \sigma^2_x \sigma^2_y +\frac{1}{8} (2 \lambda_1
+\lambda_2)\sigma^4_x + \frac{1}{4} (\lambda_1+\lambda_2)\sigma^4_y. \label{Upotio}
\end{eqnarray}

%
Here, $m^2$, $h_x$, $h_y$, $\lambda_1$, $\lambda_2$ and $c$ are model parameters as reported in Ref.~\cite{Schaefer:2008hk}. The parameters values used in the current study, are listed in Table.~\ref{par_tab1} below. Different studies~\cite{Kovacs:2015pha,Kovacs:2016juc} indicate that extending the PQM model with the vector meson sector will help to accomplish better agreement with LQCD at $T<T_c$. Such correction was not included in this work and shall be discussed in future work.

The third term in Eq. (\ref{potential}) $ \Omega_{\bar{\psi}\psi}(T,\mu_f;\phi,\phi^{*}) $ which gives the quark and anti-quark contributions can be shown as \cite{Schaefer:2008ax},
%
\begin{eqnarray} \label{z_MF}
\Omega_{\bar{\psi}\psi} (T,\mu_f;\phi,\phi^{*}) &=& -2 T \sum_{f} \int \dfrac{d^3 p}{(2 \pi)^3}
\\ \nn &&
\left\{ \ln \left[ 1+3\left(\phi+\phi^* e^{-\frac{(E_{f} - \mu_f)}{T}}\right)\, e^{-\frac{(E_{f} - \mu_f)}{T}}+e^{-3 \frac{(E_{f} - \mu_f)}{T}}\right] \right.
\\ \nn &&
\left. +\ln \left[ 1+3\left(\phi^*+\phi e^{-\frac{(E_{f} + \mu_f)}{T}}\right)\, e^{-\frac{(E_{f} + \mu_f)}{T}}+e^{-3\frac{(E_{f} + \mu_f)}{T}}\right] \right\}.
\end{eqnarray} 
%
where $N$ gives the number of the quark flavors, $E_{f}=\sqrt{\vec{p}^2+m_{f}^{2}}$ (the index $ f $ runs over different quark flavors ($u,~d~and~s$)) is the dispersion relation, energy, of the valence quark and antiquark. Assuming degenerate light quarks, $q\equiv u, d$, then we can give the masses as follows:
\begin{eqnarray}
m_q &=& g \frac{\sigma_x}{2}, \label{qmass} \\
m_s &=& g \frac{\sigma_y}{\sqrt{2}}.  \label{sqmass}
\end{eqnarray} 

The quark chemical potentials $\mu_{f}$ are related to the baryon ($\mu_B$), strange ($\mu_S$) and charge ($\mu_Q$) chemical potentials via the following transformations \cite{Fu:2013ica};
\begin{eqnarray}
\mu_{u} &=& \frac{\mu_B}{3} + \frac{2 \mu_Q}{3},\nn\\
\mu_d &=& \frac{\mu_B}{3} - \frac{\mu_Q}{3},\nn\\
\mu_s &=& \frac{\mu_B}{3} - \frac{\mu_Q}{3} - \mu_S,\nn
\label{eq.muqtomuH}
\end{eqnarray}

The influences of a finite volume are introduced in the PQM model by following the approximate method illustrated in~\cite{Bhattacharyya:2015zka,Bhattacharyya:2012rp} via a lower momentum cut-off $p_{min}[GeV]=\pi/R[GeV]=\lambda$, where $R$ is the length of a cubic volume. In this analysis, we are studying a simple situation (lower momentum cut-off ). A full implementation of the finite volume would require decent consideration of the effects of the surface and curvature, as well as boundary conditions which are periodic for bosons and anti-periodic for fermions. This full implementation of the boundary conditions leads to an infinite sum over discrete momentum values.

The PQM model has a set of parameters discussed in Refs \cite{Schaefer:2008hk,Ghosh:2007wy} and listed in Tabs.(\ref{par_tab1},\ref{par_tab2}).
%

\begin{center}
\begin{table}[hbt]
\begin{center}
 \begin{tabular}{|c|c|c|c|c|c|}
 \hline
 $c$ (MeV)& $\lambda_{1} $ & $m^{2} $ ($MeV^{2}$) & $\lambda_{2} $ & $h_{x} $ ($MeV^{3}$) & $h_{y} $ ($MeV^{3}$) \\
 \hline
 $4807.84$ & $1.40$ & $ (342.52)^2$ & $46.48$ & $(120.73)^3$ & $(336.41)^3$ \\
 \hline
\end{tabular} 
\caption{Summary of the QM model parameters employed in the presented calculations. 
\label{par_tab1}}
\end{center}
\end{table}
\end{center}

\begin{center}
\begin{table}[hbt]
\begin{center}
 \begin{tabular}{|c|c|c|c|c|c|}
 \hline
  $a_{1}$ & $a_{2}$ & $b_{0}$ & $b_{1}$& $b_{2}$& $b_{3}$ \\
 \hline
 $0.75$ & $7.5$ & $6.75$ & $-1.95$ & $2.625$ & $-7.44$ \\
 \hline
  $k_{1}$& $k_{2}$& $k_{3}$ & $k_{4}$ & $ $ & $ $ \\
 \hline
 $0.30$ & $0.25$ & $0.24$ & $0.20$ &$ $ &$ $ \\
 \hline
\end{tabular} 
\caption{Summary of the Polyakov loop potential parameters employed in the presented calculations. 
\label{par_tab2}}
\end{center}
\end{table}
\end{center}

In order to estimate the model different parameters, $\sigma_x$, $ \sigma_y$, $\phi$ and $\phi^*$,  we minimize the thermodynamic potential, Eq. (\ref{potential}), with respect to $\sigma_x$, $ \sigma_y$, $\phi$ and $\phi^*$ which gives us a set of four equations of motion:
\begin{eqnarray}\label{cond1}
\left.\frac{\partial \Omega}{\partial \sigma_x}= \frac{\partial
\Omega}{\partial \sigma_y}= \frac{\partial \Omega}{\partial
\phi}= \frac{\partial \Omega}{\partial \phi^*}\right|_{min} =0,
\end{eqnarray}
where $\sigma_x=\bar{\sigma_x}$, $\sigma_y=\bar{\sigma_y}$, $\phi=\bar{\phi}$ and $\phi^*=\bar{\phi^*}$ are the global minimum.

\section{Results}\label{sec:III}

In this section, we will discuss our PQM model calculations using the parameters summarized in Tables.~(\ref{par_tab1},\ref{par_tab2}) to illustrate the effect of finite volume on the model order parameters, phase transition and fluctuations/correlations of the conserved charges.

\subsection{Order parameters and phase transation}\label{subsec:III:I}

In the following, we present several calculations to illustrate the impacts of the finite volumes on the PQM model order parameters and the chiral phase transition. 

\begin{figure}[h!]
\centering{
\includegraphics[width=0.8\linewidth, angle=0]{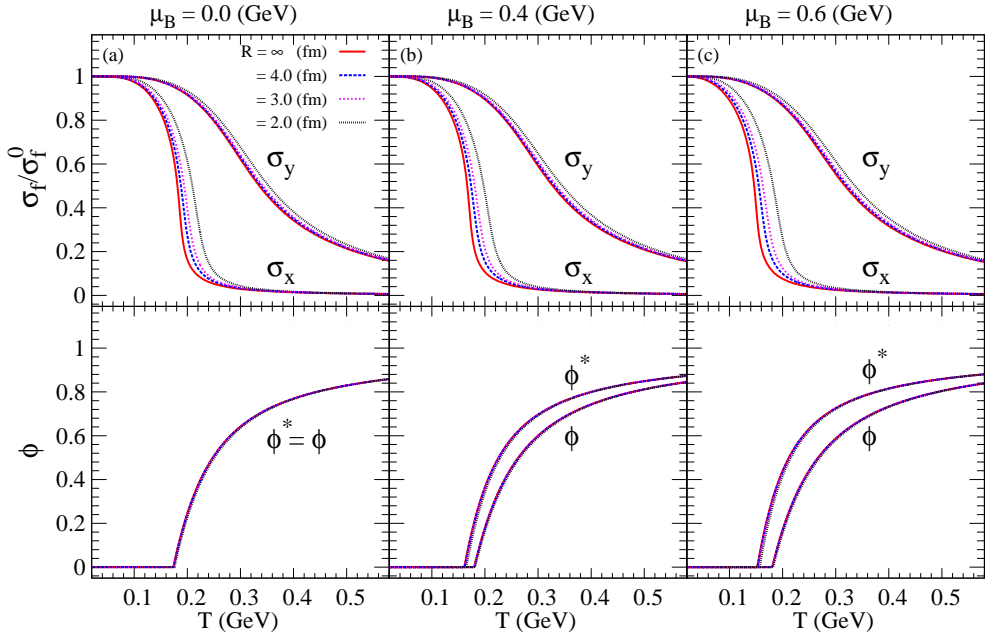}
\caption{
Temperature dependence of the (non)strange chiral condensates ($\sigma_{x}$) $\sigma_{y}$ panels (a, b and c) and the two Polyakov loops ($\phi$, $\phi^*$) panels (d, e and f) for several volume selections with $\mu_{B}$ = 0.0, 0.4 and 0.6 GeV.
\label{fig:Fig1}}}
\end{figure}

\begin{figure}[h!]
\centering{
\includegraphics[width=0.8\linewidth, angle=0]{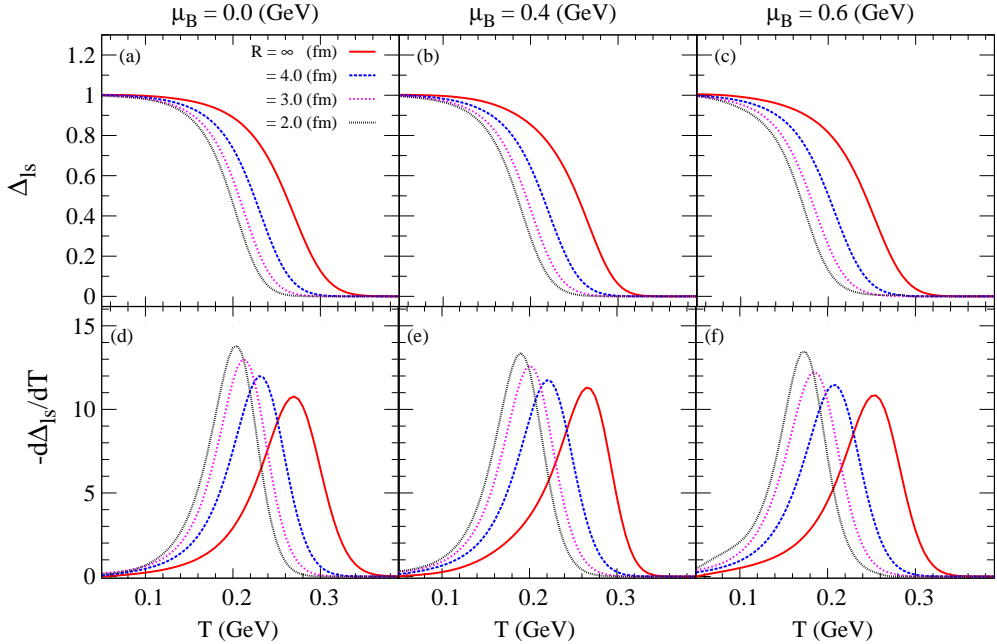}
\caption{
The same as in Fig.(\ref{fig:Fig1}) but for the net-difference condensate ($\Delta_{ls}$) panels (a, b and c) and $d\Delta_{ls}/dT$ panels (d, e and f). \label{fig:Fig2}
}}
\end{figure}

\begin{figure}[h!]
\centering{
\includegraphics[width=0.4\linewidth, angle=-90]{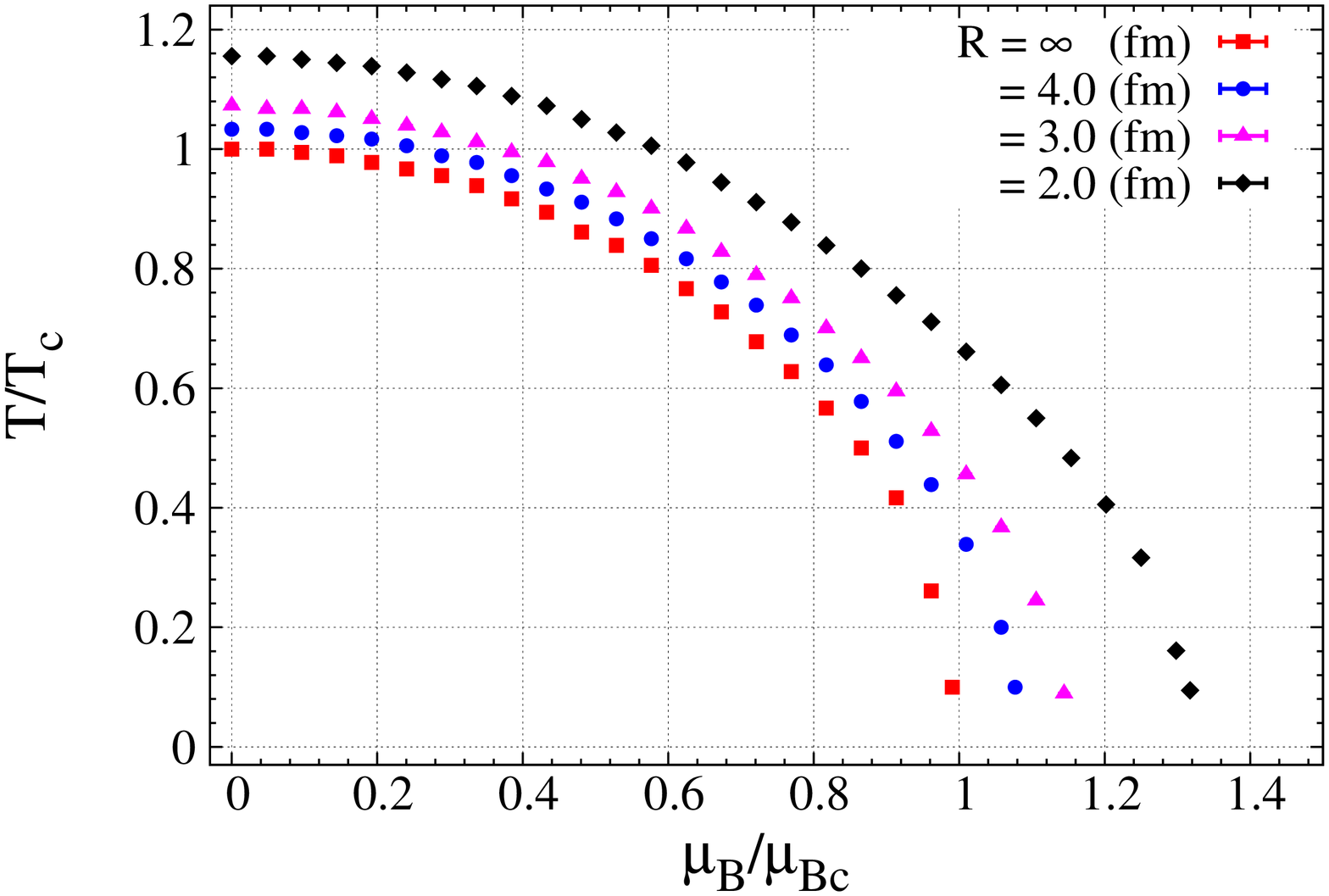}
\caption{
Chiral phase diagram for different volumes selections.
\label{fig:Fig3}}}
\end{figure}

Figure.~\ref{fig:Fig1} shows the thermal dependence of the non-strange and strange chiral condensates ($\sigma_{x}$, $\sigma_{y}$) panels (a, b and c) and the Polyakov loops ($\phi$ and $\phi^{*}$) panels (d, e and f) for different volume  selections and different $\mu_{B}$ values. The upper panels show that both $\sigma_{x}$ and $\sigma_{y}$ increase as the system volume is decreased, with larger sensitivity for the non-strange chiral condensates ($\sigma_{x}$). The lower panels show very little if any, volume dependence for $\phi$ and $\phi^{*}$ at different $\mu_{B}$. 

\begin{figure}[h!]
\centering{
\includegraphics[width=0.4\linewidth, angle=-90]{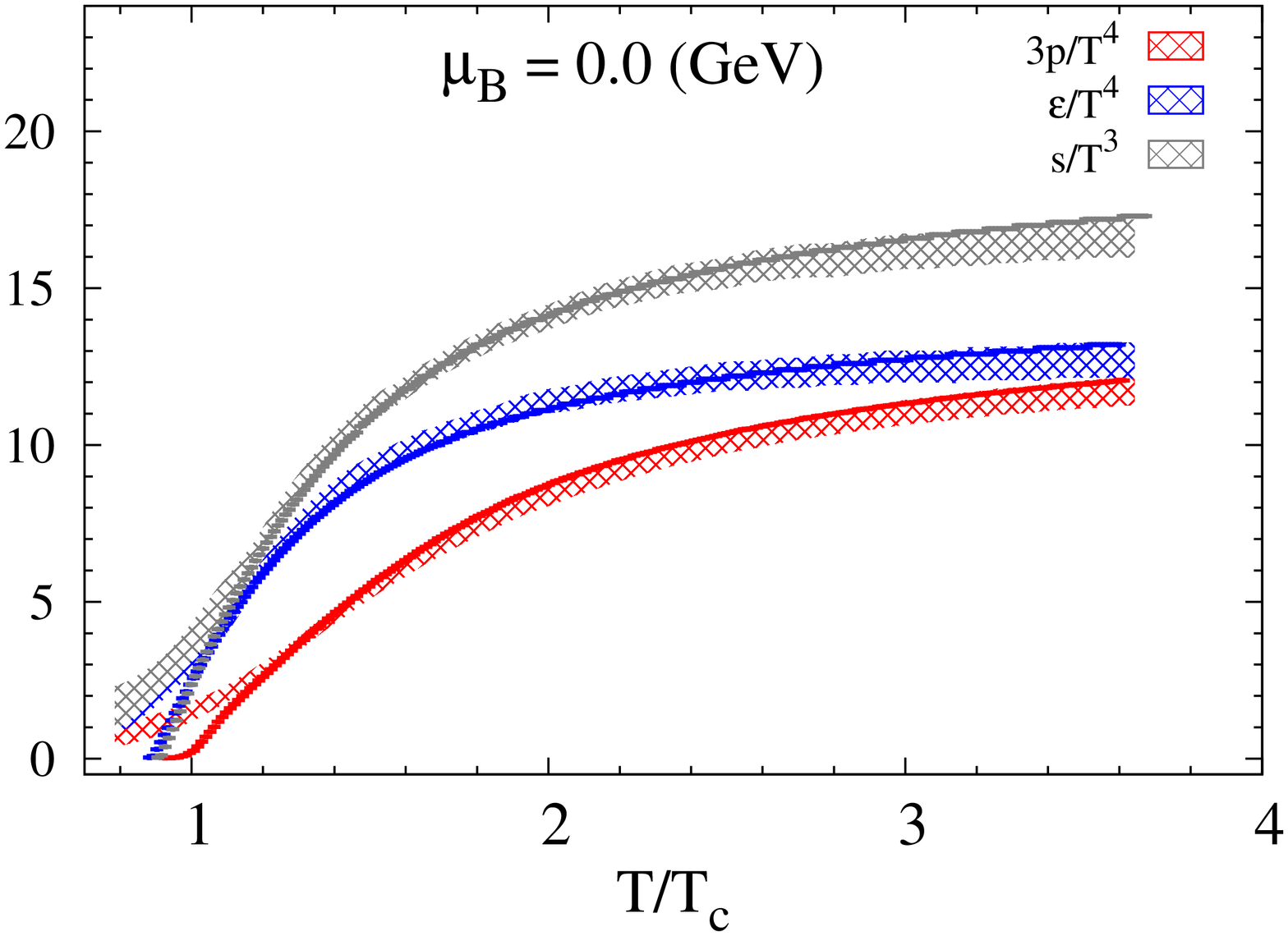}
\caption{
Comparison of the PQM model pressure density, energy density and entropy to results from LQCD. The comparisons are made for $\mu_{B}$ = 0.0 GeV; the lines indicate the PQM model calculations and the shaded areas indicate LQCD results from Ref.~\cite{Borsanyi:2013bia}.
\label{fig:Fig4}}}
\end{figure}

\begin{figure}[h!]
\centering{
\includegraphics[width=5.cm,angle=-90]{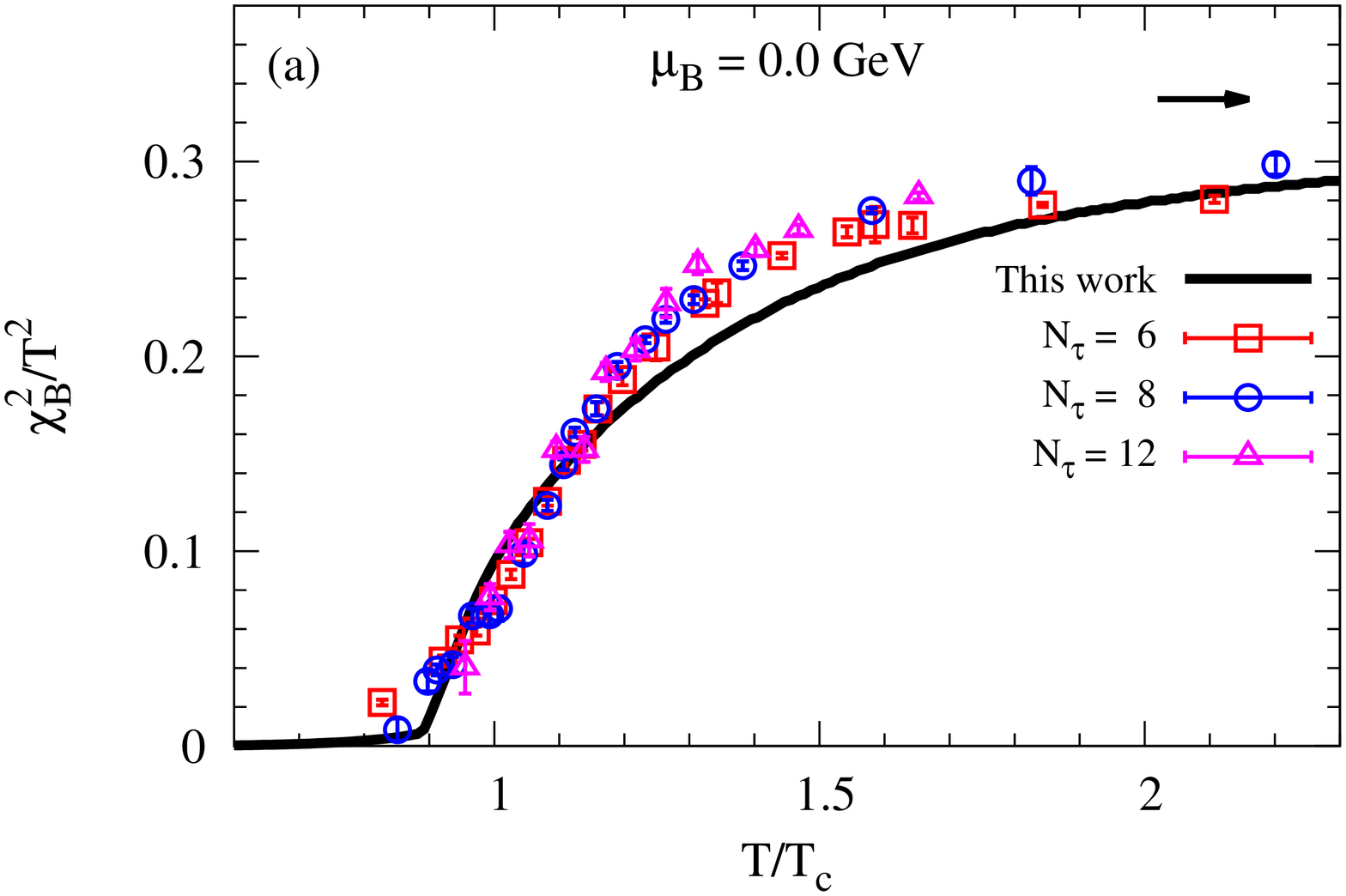}
\includegraphics[width=5.cm,angle=-90]{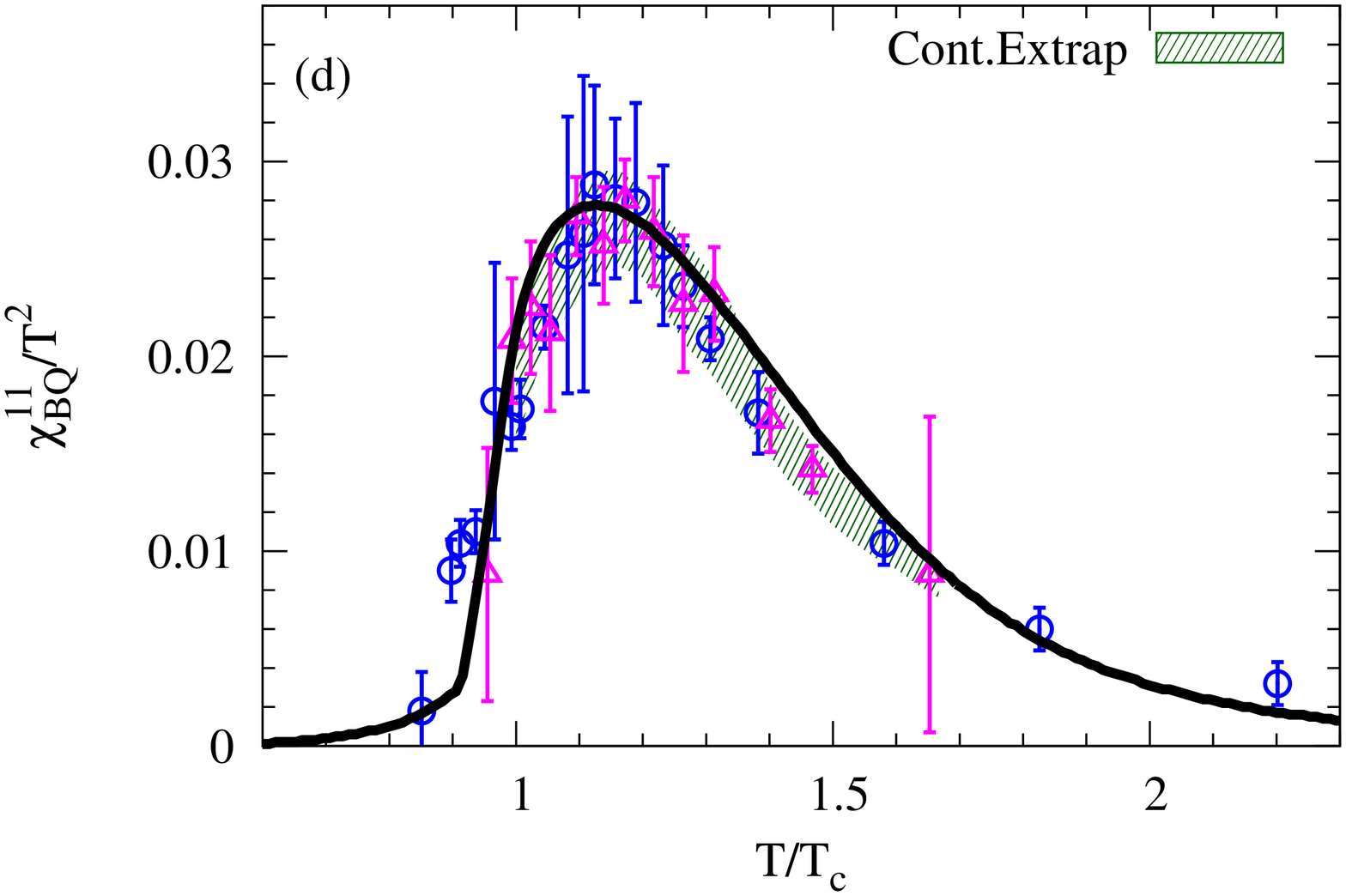}
\includegraphics[width=5.cm,angle=-90]{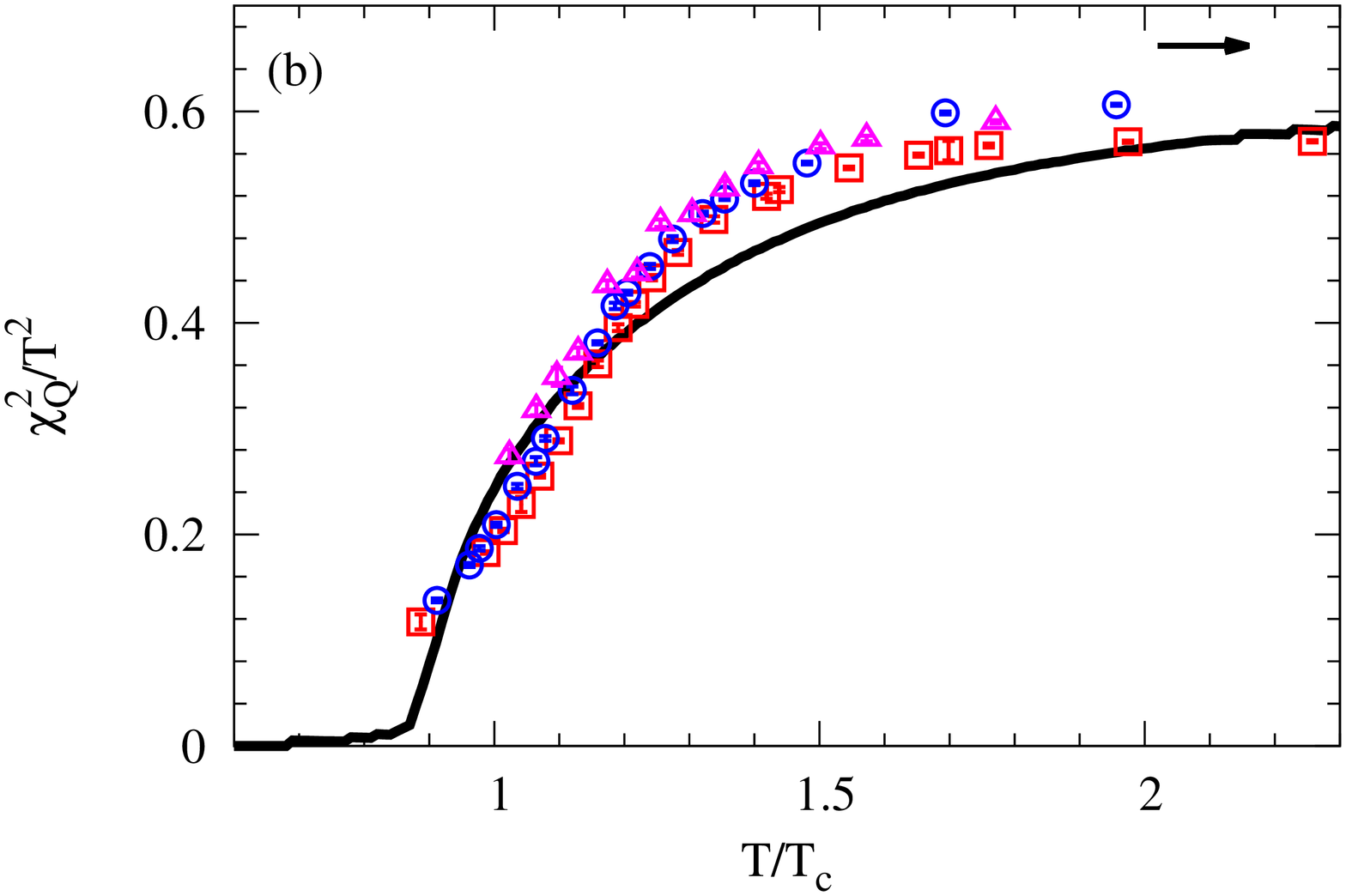}
\includegraphics[width=5.cm,angle=-90]{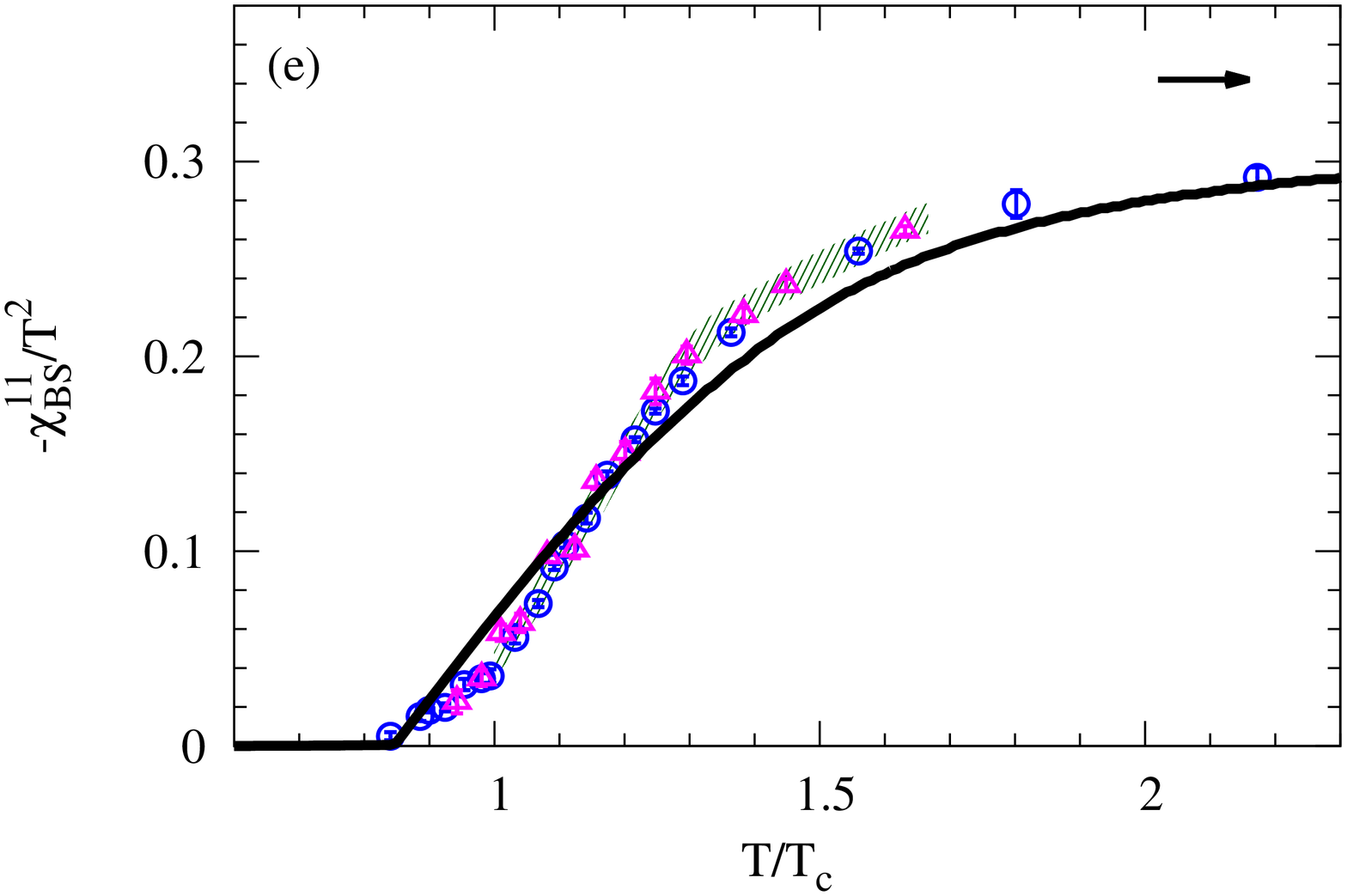}
\includegraphics[width=5.cm,angle=-90]{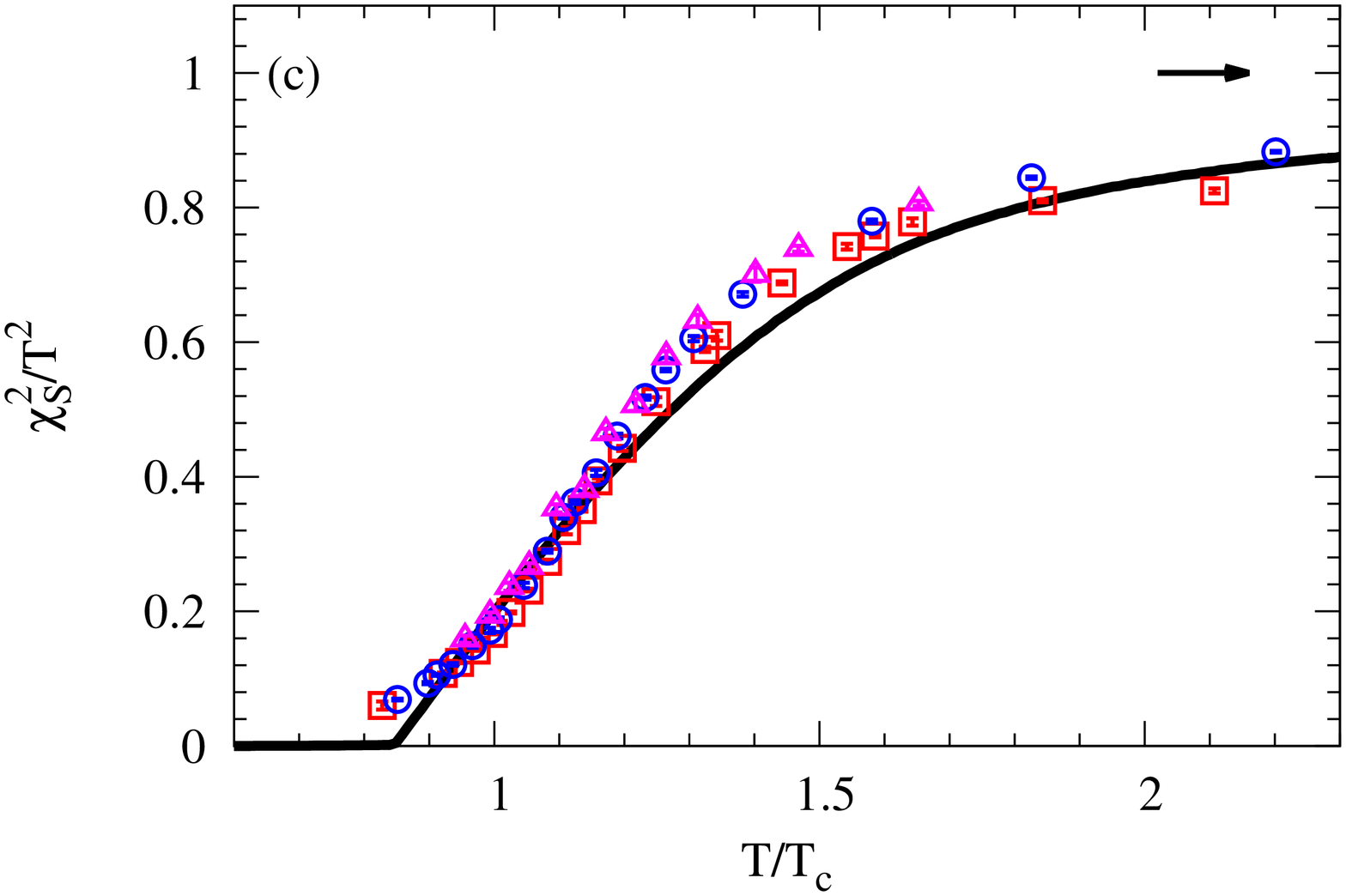}
\includegraphics[width=5.cm,angle=-90]{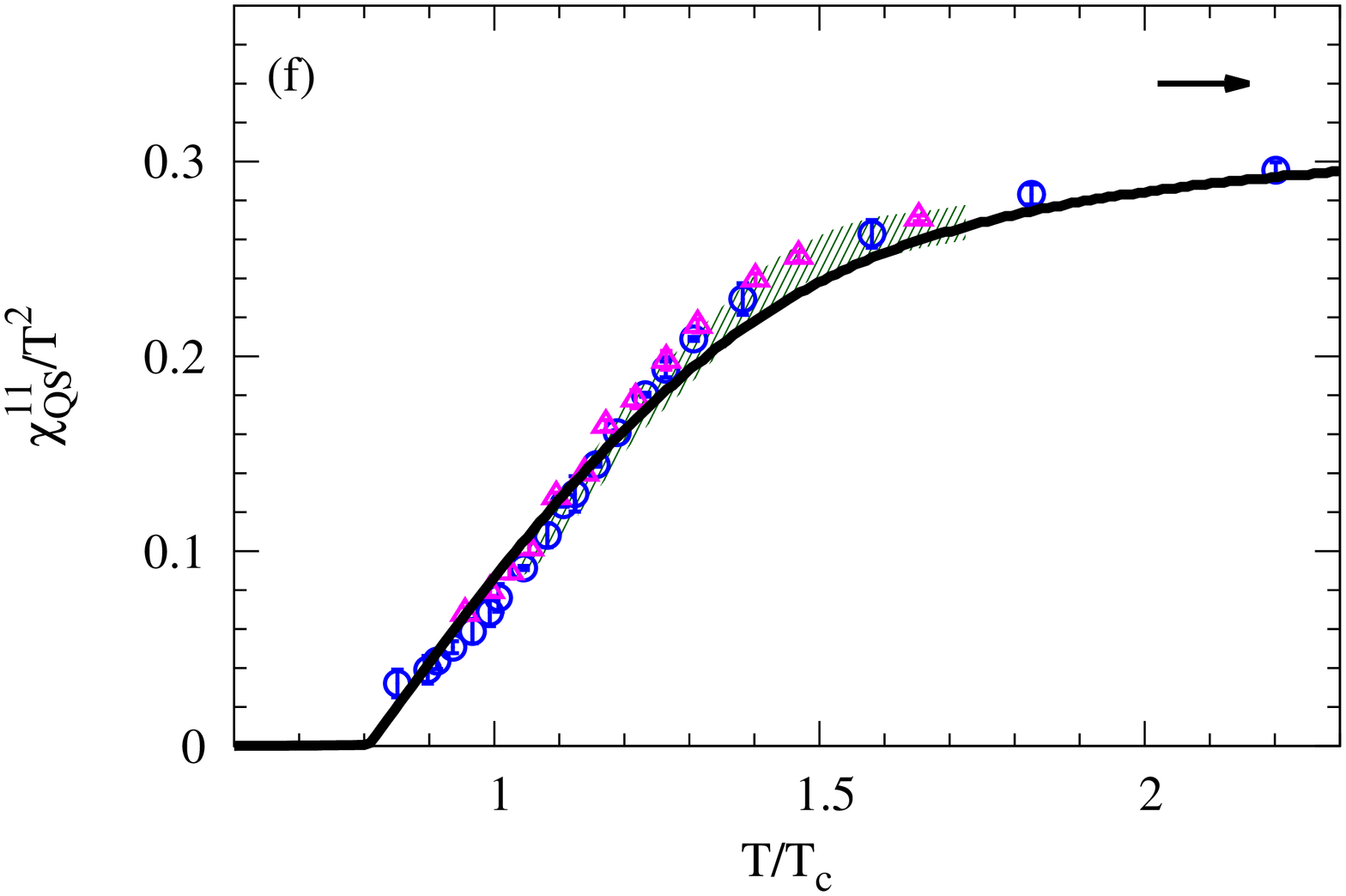}
\caption{The thermal behavior of PQM model conserved charges fluctuation (a, b and c) and correlation (d, e and f) are compared with the same quantities obtained from the LQCD (symbols). The comparisons are made for $\mu_B = 0$. The LQCD conserved charges fluctuation (a, b and c) and correlation (d, e and f) are taken from Ref~\cite{Bazavov:2012jq} tables (III, IV and V).
 \label{fig:Fig5}}} 
\end{figure}

As pointed out PQM model contains strange and non-strange chiral condensates which reflect the chiral phase transitions. Using both chiral condensates we can investigate the finite volume effects on the $SU(3)$ $2+1$ PQM model chiral phase transition via the normalized net-difference condensate $\Delta_{ls}(T)$ as defined in Ref.~\cite{Schaefer:2009ui},
\begin{eqnarray}
 \Delta_ {ls}(T) &=& \dfrac{\sigma_x - \dfrac{h_x}{h_y} \sigma_y}{\sigma_{x0} - \dfrac{h_x}{h_y} \sigma_{y0}},\label{Eq:Delta}
\end{eqnarray}
where $h_x$ ($h_y$) are non-strange (strange) explicit symmetry breaking parameters. 

Figure.~\ref{fig:Fig2} shows the thermal dependence of the normalized net-difference condensate $\Delta_{ls}$ panels (a, b and c) and $d\Delta_{ls}/dT$ panels (d, e and f) for different volume selections and different $\mu_{B}$ values. The upper panels indicate an increase in $\Delta_{ls}(T)$ as the system volume is decreased. The lower panels show that for fixed values of $R$ and $\mu_B$, the $d\Delta_{ls}/dT$ is peaking up at a specific point indicating the phase transition. The peak position is shifted toward lower temperature as the $\mu_{B}$ value increase.
The study of the phase diagram of the PQM model for at fixed volume could be done through mapping out the $\mu_{B}$ dependence of $\Delta_{q,s}(T)$. For a fixed $R$ and $\mu_B$ values, $d\Delta_{ls}/dT$ will peak up at a particular point expressing the phase transition. Therefore, the phase diagram can be studied by outlining such points for a wide range of baryon chemical potentials. Fig. \ref{fig:Fig3} illustrates the effects of finite volume on the phase diagram. The parameters $T_{c}$ and $\mu_{Bc}$ represent the transition temperature at $\mu_{B}$ = 0.0 GeV and the transition chemical potential at low temperature respectively at $R$ = $\infty$. Our calculations reveal that the PQM model phase diagram in the $(\mu_{B},T)$-plane, increases with decreasing the system volume. For the $R$ = 2.0 (fm) the  $\mu_{B}$ value at low temperature increased by about $30\%$ and the $T$ value at $\mu_{B}$ = 0.0 GeV  increased by about $19\%$ from them values at $R$ = $\infty$ (fm).

\subsection{Fluctuations and correlations of conserved charges}\label{subsec:III:II}
%
The thermodynamics quantities and (diagonal) off-diagonal susceptibilities can be determined by using the thermodynamic pressure as \cite{Bazavov:2012jq},

\begin{figure}[h!]
\centering{
\includegraphics[width=0.75\linewidth, angle=0]{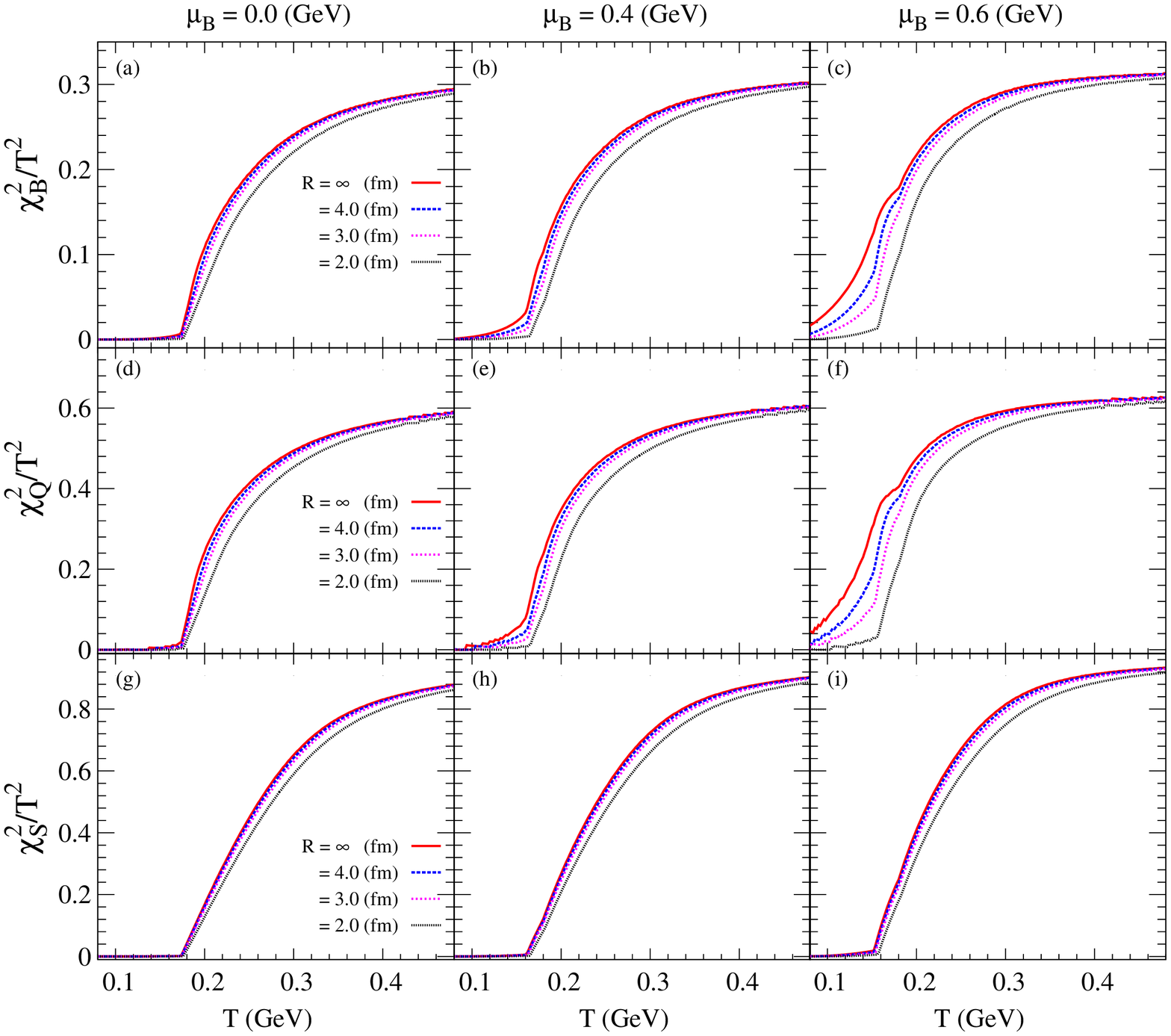}
\caption{
The thermal behavior of the normalized diagonal susceptibilities, $\chi^{2}_{BB}$, $\chi^{2}_{QQ}$ and $\chi^{2}_{SS}$, for several volume selections at $\mu_B$ = 0.0, 0.4 and 0.6 GeV. 
\label{fig:Fig6}}}
\end{figure}

\begin{figure}[h!]
\centering{
\includegraphics[width=0.75\linewidth, angle=0]{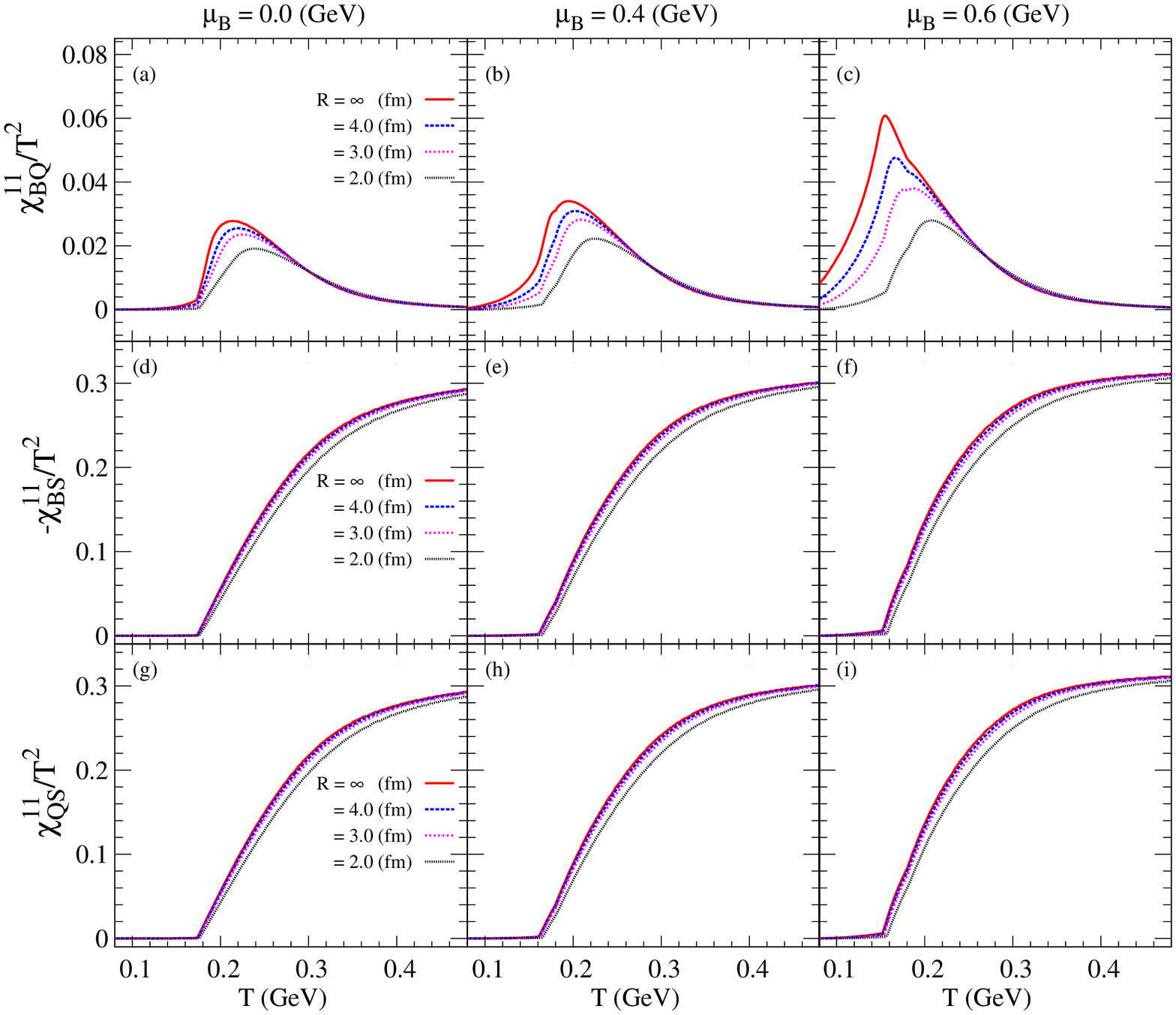}
\caption{
The thermal behavior of the normalized off-diagonal susceptibilities, $\chi^{11}_{BQ}$, $\chi^{11}_{BS}$ and $\chi^{11}_{QS}$, for several volume selections at $\mu_B$ = 0.0, 0.4 and 0.6 GeV. 
\label{fig:Fig7}}}
\end{figure}

\begin{figure}[h!]
\centering{
\includegraphics[width=0.75\linewidth, angle=0]{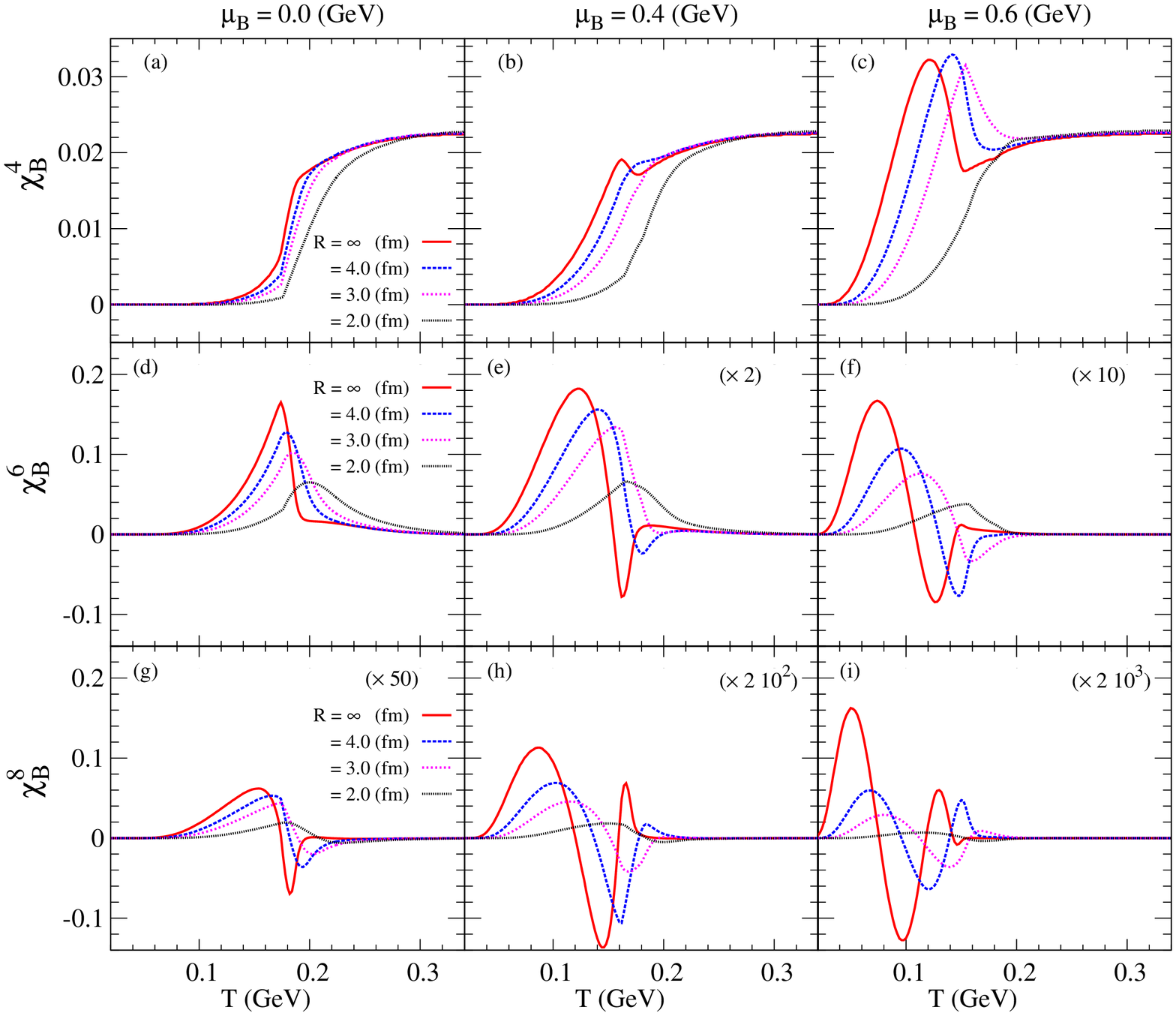}
\caption{
The thermal behavior of the higher order susceptibilities, $\chi^{4}_{B}$, $\chi^{6}_{B}$ and $\chi^{8}_{B}$, for several volume selections at $\mu_B$ = 0.0, 0.4 and 0.6 GeV. 
\label{fig:Fig8}}}
\end{figure}

\begin{eqnarray}
p &=& - \Omega(T, \mu_f),\\
s &=&   dp/dT,\\
\epsilon &=&  T s - p,\\
\end{eqnarray}

\begin{eqnarray}
\chi^{ijk}_{BQS} &=& \frac{\partial^{i+j+k}(p)}{\partial(\mu_{B})^{i} \partial(\mu_{Q})^{j} \partial(\mu_{S})^{k}}. \label{eq.sus}
\end{eqnarray}
where superscripts $i$, $j$ and $k$ run over integers that represent the derivatives orders. The indexes $B$, $Q$ and $S$ represent the conserved-quantities, baryon, charge, and strangeness, respectively. Eq. (\ref{eq.sus}) illustrates the dependence of the (fluctuations) correlations of conserved charges on the temperature, chemical potential, and system volume.  The susceptibilities evaluated first by computing the thermodynamic potential at vanishing $\mu_{f}$ and then expand the scaled  thermodynamic potential in a Taylor series around $\mu_{f}/T = 0$.

Before addressing the system volume effect, it is informative to contrast the PQM model thermodynamics quantities and (diagonal) off-diagonal susceptibilities calculations for ($\mu_B = 0$ and $R = \infty$), to similar results from LQCD calculations~\cite{Bazavov:2012jq,Borsanyi:2013bia}. Such comparisons are presented in Figs.~(\ref{fig:Fig4},\ref{fig:Fig5}); which indicate a good agreement between the PQM model and LQCD~\cite{Bazavov:2012jq,Borsanyi:2013bia}. These comparisons could be improved spatially at low temperature by including the vector mesons sector to the PQM model. The influence of the finite volume on the model thermodynamics quantities has been discussed in our previous study~\cite{Magdy:2015eda}.

Figure~\ref{fig:Fig6} displays the temperature dependence of the normalized conserved-fluctuations, baryon ($\chi_{BB}$), charge ($\chi_{QQ}$) and strangeness ($\chi_{SS}$), respectively. The results presented for several volume selections at three $\mu_B$ values, $\mu_B = 0.0,~0.4~and~0.6$~GeV. Our results indicate that the normalized fluctuations decrease with the volume which quickly trends towards the infinite volume value at high temperature. The non-strange susceptibilities ($\chi^{2}_{BB}$ and $\chi^{2}_{QQ}$) shows a higher sensitivity to the volume change more than the strange susceptibility ($\chi^{2}_{SS}$). This weak sensitivity to the volume change of the strange quantities could be driving from the large mass of the strange quark.

Figure~\ref{fig:Fig7} shows the temperature dependence of the off-diagonal susceptibilities, $\chi^{2}_{BQ}$, $\chi^{2}_{BS}$ and $\chi^{2}_{SQ}$ for several volume selections and different $\mu_B$ values. The net baryons show a high correlation to the net charge and less correlation to the net strange. Our results indicate that the normalized correlations decrease with the volume which quickly trends towards the infinite volume value at high temperature. Also, the non-strange correlation ($\chi_{BQ}$) show a higher sensitivity to the volume change.

Also, the temperature dependence of the higher order baryon susceptibilities $\chi^{n}_{B}$ ($n$ = 4, 6 and 8) for different volume selections at $\mu_B$ values, $\mu_B$ = 0.0, 0.4 and 0.6 GeV are shown in Figure~\ref{fig:Fig8}. The $n^{th}$-order susceptibilities decrease with the volume selections, and for $n = 6, 8$ they start to peak around the transition temperature $T_{c}$. Also, we observed a stronger oscillation in all higher harmonics as we increase the $\mu_B$ values.

\section{Conclusions}\label{sec:IIII}
In this work, we have used the $2+1$ $SU(3)$ Polyakov Quark-Meson model (PQM) framework to study the properties of the QCD medium produced at finite volume in heavy ion collisions. This model framework provides several conserved-quantities, baryon, charge, and strangeness which compare well with those obtained in LQCD calculations for vanishing $\mu_B$. 
Our calculations indicate that the conserved-quantities ($\chi^{ijk}_{BQS}$) are significantly influenced by finite volume effects. The calculated conserved-quantities decrease with the volume which quickly trends towards the infinite volume value at high temperature. Also, the non-strange quantities show a higher sensitivity to the volume change more than the strange once. Finally, PQM model conserved-quantities, suggests that the quark-hadron phase boundary is shifted to higher values of $\mu_{B}$ and $T$ with decreasing system volume. 

\section*{Acknowledgements}
This work is supported by the US Department of Energy under contract DE-FG02-94ER40865.

%
%
\bibliography{PLSM_volum} 

\begin{thebibliography}{38}%
\makeatletter
\providecommand \@ifxundefined [1]{%
 \@ifx{#1\undefined}
}%
\providecommand \@ifnum [1]{%
 \ifnum #1\expandafter \@firstoftwo
 \else \expandafter \@secondoftwo
 \fi
}%
\providecommand \@ifx [1]{%
 \ifx #1\expandafter \@firstoftwo
 \else \expandafter \@secondoftwo
 \fi
}%
\providecommand \natexlab [1]{#1}%
\providecommand \enquote  [1]{``#1''}%
\providecommand \bibnamefont  [1]{#1}%
\providecommand \bibfnamefont [1]{#1}%
\providecommand \citenamefont [1]{#1}%
\providecommand \href@noop [0]{\@secondoftwo}%
\providecommand \href [0]{\begingroup \@sanitize@url \@href}%
\providecommand \@href[1]{\@@startlink{#1}\@@href}%
\providecommand \@@href[1]{\endgroup#1\@@endlink}%
\providecommand \@sanitize@url [0]{\catcode `\\12\catcode `\$12\catcode
  `\&12\catcode `\#12\catcode `\^12\catcode `\_12\catcode `\%12\relax}%
\providecommand \@@startlink[1]{}%
\providecommand \@@endlink[0]{}%
\providecommand \url  [0]{\begingroup\@sanitize@url \@url }%
\providecommand \@url [1]{\endgroup\@href {#1}{\urlprefix }}%
\providecommand \urlprefix  [0]{URL }%
\providecommand \Eprint [0]{\href }%
\providecommand \doibase [0]{http://dx.doi.org/}%
\providecommand \selectlanguage [0]{\@gobble}%
\providecommand \bibinfo  [0]{\@secondoftwo}%
\providecommand \bibfield  [0]{\@secondoftwo}%
\providecommand \translation [1]{[#1]}%
\providecommand \BibitemOpen [0]{}%
\providecommand \bibitemStop [0]{}%
\providecommand \bibitemNoStop [0]{.\EOS\space}%
\providecommand \EOS [0]{\spacefactor3000\relax}%
\providecommand \BibitemShut  [1]{\csname bibitem#1\endcsname}%
\let\auto@bib@innerbib\@empty
\bibitem [{\citenamefont {Fachini}(2006)}]{Fachini:2006ch}%
  \BibitemOpen
  \bibfield  {author} {\bibinfo {author} {\bibfnamefont {P.}~\bibnamefont
  {Fachini}},\ }\bibfield  {title} {\enquote {\bibinfo {title} {{Experimental
  highlights of the RHIC program}},}\ }\bibfield  {booktitle} {\emph {\bibinfo
  {booktitle} {{Proceedings, 10th Mexican Workshop on Particles and Fields
  (MWPF 2005), Part A and B: Morelia, Mexico, November 6-12, 2005}}},\ }\href
  {\doibase 10.1063/1.2359243} {\bibfield  {journal} {\bibinfo  {journal} {AIP
  Conf. Proc.}\ }\textbf {\bibinfo {volume} {857}},\ \bibinfo {pages} {62--75}
  (\bibinfo {year} {2006})},\ \Eprint {http://arxiv.org/abs/hep-ex/0605102}
  {arXiv:hep-ex/0605102 [hep-ex]} \BibitemShut {NoStop}%
\bibitem [{\citenamefont {Monteno}(2007)}]{Monteno:2007gi}%
  \BibitemOpen
  \bibfield  {author} {\bibinfo {author} {\bibfnamefont {M.}~\bibnamefont
  {Monteno}} (\bibinfo {collaboration} {ALICE}),\ }\bibfield  {title} {\enquote
  {\bibinfo {title} {{The physics programme of the ALICE experiment at the
  LHC}},}\ }\bibfield  {booktitle} {\emph {\bibinfo {booktitle} {{Perspectives
  in hadronic physics. Proceedings, 5th International Conference on
  perspectives in hadronic physics, particle-nucleus and nucleus-nucleus
  scattering at relativistic energies, Trieste, Italy, May 22-26, 2006}}},\
  }\href {\doibase 10.1016/j.nuclphysa.2006.10.012} {\bibfield  {journal}
  {\bibinfo  {journal} {Nucl. Phys.}\ }\textbf {\bibinfo {volume} {A782}},\
  \bibinfo {pages} {283--290} (\bibinfo {year} {2007})}\BibitemShut {NoStop}%
\bibitem [{\citenamefont {Aoki}\ \emph {et~al.}(2006)\citenamefont {Aoki},
  \citenamefont {Endrodi}, \citenamefont {Fodor}, \citenamefont {Katz},\ and\
  \citenamefont {Szabo}}]{Aoki:2006we}%
  \BibitemOpen
  \bibfield  {author} {\bibinfo {author} {\bibfnamefont {Y.}~\bibnamefont
  {Aoki}}, \bibinfo {author} {\bibfnamefont {G.}~\bibnamefont {Endrodi}},
  \bibinfo {author} {\bibfnamefont {Z.}~\bibnamefont {Fodor}}, \bibinfo
  {author} {\bibfnamefont {S.~D.}\ \bibnamefont {Katz}}, \ and\ \bibinfo
  {author} {\bibfnamefont {K.~K.}\ \bibnamefont {Szabo}},\ }\bibfield  {title}
  {\enquote {\bibinfo {title} {{The Order of the quantum chromodynamics
  transition predicted by the standard model of particle physics}},}\ }\href
  {\doibase 10.1038/nature05120} {\bibfield  {journal} {\bibinfo  {journal}
  {Nature}\ }\textbf {\bibinfo {volume} {443}},\ \bibinfo {pages} {675--678}
  (\bibinfo {year} {2006})},\ \Eprint {http://arxiv.org/abs/hep-lat/0611014}
  {arXiv:hep-lat/0611014 [hep-lat]} \BibitemShut {NoStop}%
\bibitem [{\citenamefont {Ejiri}(2008)}]{Ejiri:2008xt}%
  \BibitemOpen
  \bibfield  {author} {\bibinfo {author} {\bibfnamefont {Shinji}\ \bibnamefont
  {Ejiri}},\ }\bibfield  {title} {\enquote {\bibinfo {title} {{Canonical
  partition function and finite density phase transition in lattice QCD}},}\
  }\href {\doibase 10.1103/PhysRevD.78.074507} {\bibfield  {journal} {\bibinfo
  {journal} {Phys. Rev.}\ }\textbf {\bibinfo {volume} {D78}},\ \bibinfo {pages}
  {074507} (\bibinfo {year} {2008})},\ \Eprint {http://arxiv.org/abs/0804.3227}
  {arXiv:0804.3227 [hep-lat]} \BibitemShut {NoStop}%
\bibitem [{\citenamefont {Pisarski}\ and\ \citenamefont
  {Wilczek}(1984)}]{Pisarski:1983ms}%
  \BibitemOpen
  \bibfield  {author} {\bibinfo {author} {\bibfnamefont {Robert~D.}\
  \bibnamefont {Pisarski}}\ and\ \bibinfo {author} {\bibfnamefont {Frank}\
  \bibnamefont {Wilczek}},\ }\bibfield  {title} {\enquote {\bibinfo {title}
  {{Remarks on the Chiral Phase Transition in Chromodynamics}},}\ }\href
  {\doibase 10.1103/PhysRevD.29.338} {\bibfield  {journal} {\bibinfo  {journal}
  {Phys. Rev.}\ }\textbf {\bibinfo {volume} {D29}},\ \bibinfo {pages}
  {338--341} (\bibinfo {year} {1984})}\BibitemShut {NoStop}%
\bibitem [{\citenamefont {Lee}(1972)}]{Chiral:dynamics}%
  \BibitemOpen
  \bibfield  {author} {\bibinfo {author} {\bibfnamefont {Benjamin~W}\
  \bibnamefont {Lee}},\ }\bibfield  {title} {\enquote {\bibinfo {title}
  {{Chiral dynamics}},}\ }\href@noop {} {\bibfield  {journal} {\bibinfo
  {journal} {New York, NY : Gordon and Breach,}\ }\textbf {\bibinfo {volume}
  {B591}},\ \bibinfo {pages} {129 p.} (\bibinfo {year} {1972})}\BibitemShut
  {NoStop}%
\bibitem [{\citenamefont {Kovacs}\ and\ \citenamefont
  {Szep}(2007)}]{Kovacs:2006ym}%
  \BibitemOpen
  \bibfield  {author} {\bibinfo {author} {\bibfnamefont {P.}~\bibnamefont
  {Kovacs}}\ and\ \bibinfo {author} {\bibfnamefont {Zs.}\ \bibnamefont
  {Szep}},\ }\bibfield  {title} {\enquote {\bibinfo {title} {{The critical
  surface of the $SU(3)_L$ $x$ $SU(3)_R$ chiral quark model at non-zero baryon
  density}},}\ }\href {\doibase 10.1103/PhysRevD.75.025015} {\bibfield
  {journal} {\bibinfo  {journal} {Phys. Rev.}\ }\textbf {\bibinfo {volume}
  {D75}},\ \bibinfo {pages} {025015} (\bibinfo {year} {2007})},\ \Eprint
  {http://arxiv.org/abs/hep-ph/0611208} {arXiv:hep-ph/0611208 [hep-ph]}
  \BibitemShut {NoStop}%
\bibitem [{\citenamefont {Kovacs}\ and\ \citenamefont
  {Szep}(2008)}]{Kovacs:2007sy}%
  \BibitemOpen
  \bibfield  {author} {\bibinfo {author} {\bibfnamefont {P.}~\bibnamefont
  {Kovacs}}\ and\ \bibinfo {author} {\bibfnamefont {Zs.}\ \bibnamefont
  {Szep}},\ }\bibfield  {title} {\enquote {\bibinfo {title} {{Influence of the
  isospin and hypercharge chemical potentials on the location of the CEP in the
  mu(B) - T phase diagram of the $SU(3)(L)$ $x$ $SU(3)(R)$ chiral quark
  model}},}\ }\href {\doibase 10.1103/PhysRevD.77.065016} {\bibfield  {journal}
  {\bibinfo  {journal} {Phys. Rev.}\ }\textbf {\bibinfo {volume} {D77}},\
  \bibinfo {pages} {065016} (\bibinfo {year} {2008})},\ \Eprint
  {http://arxiv.org/abs/0710.1563} {arXiv:0710.1563 [hep-ph]} \BibitemShut
  {NoStop}%
\bibitem [{\citenamefont {Nambu}\ and\ \citenamefont
  {Jona-Lasinio}(1961)}]{Nambu:1961tp}%
  \BibitemOpen
  \bibfield  {author} {\bibinfo {author} {\bibfnamefont {Yoichiro}\
  \bibnamefont {Nambu}}\ and\ \bibinfo {author} {\bibfnamefont
  {G.}~\bibnamefont {Jona-Lasinio}},\ }\bibfield  {title} {\enquote {\bibinfo
  {title} {{Dynamical Model of Elementary Particles Based on an Analogy with
  Superconductivity. 1.}}}\ }\href {\doibase 10.1103/PhysRev.122.345}
  {\bibfield  {journal} {\bibinfo  {journal} {Phys. Rev.}\ }\textbf {\bibinfo
  {volume} {122}},\ \bibinfo {pages} {345--358} (\bibinfo {year} {1961})},\
  \bibinfo {note} {[,127(1961)]}\BibitemShut {NoStop}%
\bibitem [{\citenamefont {Fukushima}(2004)}]{Fukushima:2003fw}%
  \BibitemOpen
  \bibfield  {author} {\bibinfo {author} {\bibfnamefont {Kenji}\ \bibnamefont
  {Fukushima}},\ }\bibfield  {title} {\enquote {\bibinfo {title} {{Chiral
  effective model with the Polyakov loop}},}\ }\href {\doibase
  10.1016/j.physletb.2004.04.027} {\bibfield  {journal} {\bibinfo  {journal}
  {Phys. Lett.}\ }\textbf {\bibinfo {volume} {B591}},\ \bibinfo {pages}
  {277--284} (\bibinfo {year} {2004})},\ \Eprint
  {http://arxiv.org/abs/hep-ph/0310121} {arXiv:hep-ph/0310121 [hep-ph]}
  \BibitemShut {NoStop}%
\bibitem [{\citenamefont {Kahara}\ and\ \citenamefont
  {Tuominen}(2008)}]{Kahara:2008yg}%
  \BibitemOpen
  \bibfield  {author} {\bibinfo {author} {\bibfnamefont {Topi}\ \bibnamefont
  {Kahara}}\ and\ \bibinfo {author} {\bibfnamefont {Kimmo}\ \bibnamefont
  {Tuominen}},\ }\bibfield  {title} {\enquote {\bibinfo {title} {{Degrees of
  freedom and the phase transitions of two flavor QCD}},}\ }\href {\doibase
  10.1103/PhysRevD.78.034015} {\bibfield  {journal} {\bibinfo  {journal} {Phys.
  Rev.}\ }\textbf {\bibinfo {volume} {D78}},\ \bibinfo {pages} {034015}
  (\bibinfo {year} {2008})},\ \Eprint {http://arxiv.org/abs/0803.2598}
  {arXiv:0803.2598 [hep-ph]} \BibitemShut {NoStop}%
\bibitem [{\citenamefont {Wambach}\ \emph {et~al.}(2010)\citenamefont
  {Wambach}, \citenamefont {Schaefer},\ and\ \citenamefont
  {Wagner}}]{Wambach:2009ee}%
  \BibitemOpen
  \bibfield  {author} {\bibinfo {author} {\bibfnamefont {Jochen}\ \bibnamefont
  {Wambach}}, \bibinfo {author} {\bibfnamefont {Bernd-Jochen}\ \bibnamefont
  {Schaefer}}, \ and\ \bibinfo {author} {\bibfnamefont {Mathias}\ \bibnamefont
  {Wagner}},\ }\bibfield  {title} {\enquote {\bibinfo {title} {{QCD
  Thermodynamics: Confronting the Polyakov-Quark-Meson Model with Lattice
  QCD}},}\ }\bibfield  {booktitle} {\emph {\bibinfo {booktitle} {{Three days of
  strong interactions. Proceedings, EMMI Workshop and 26th Max Born Symposium,
  Wroclaw, Poland, July 9-11, 2009}}},\ }\href@noop {} {\bibfield  {journal}
  {\bibinfo  {journal} {Acta Phys. Polon. Supp.}\ }\textbf {\bibinfo {volume}
  {3}},\ \bibinfo {pages} {691--700} (\bibinfo {year} {2010})},\ \Eprint
  {http://arxiv.org/abs/0911.0296} {arXiv:0911.0296 [hep-ph]} \BibitemShut
  {NoStop}%
\bibitem [{\citenamefont {Schaefer}\ and\ \citenamefont
  {Wagner}(2009{\natexlab{a}})}]{Schaefer:2008ax}%
  \BibitemOpen
  \bibfield  {author} {\bibinfo {author} {\bibfnamefont {Bernd-Jochen}\
  \bibnamefont {Schaefer}}\ and\ \bibinfo {author} {\bibfnamefont {Mathias}\
  \bibnamefont {Wagner}},\ }\bibfield  {title} {\enquote {\bibinfo {title} {{On
  the QCD phase structure from effective models}},}\ }\bibfield  {booktitle}
  {\emph {\bibinfo {booktitle} {{Heavy-ion collisions from the Coulomb barrier
  to the quark-gluon plasma. Proceedings, International Workshop on Nuclear
  Physics, 30th Course, Erice, Italy, September 16-24, 2008}}},\ }\href
  {\doibase 10.1016/j.ppnp.2008.12.009} {\bibfield  {journal} {\bibinfo
  {journal} {Prog. Part. Nucl. Phys.}\ }\textbf {\bibinfo {volume} {62}},\
  \bibinfo {pages} {381} (\bibinfo {year} {2009}{\natexlab{a}})},\ \Eprint
  {http://arxiv.org/abs/0812.2855} {arXiv:0812.2855 [hep-ph]} \BibitemShut
  {NoStop}%
\bibitem [{\citenamefont {Mao}\ \emph {et~al.}(2010)\citenamefont {Mao},
  \citenamefont {Jin},\ and\ \citenamefont {Huang}}]{Mao:2009aq}%
  \BibitemOpen
  \bibfield  {author} {\bibinfo {author} {\bibfnamefont {Hong}\ \bibnamefont
  {Mao}}, \bibinfo {author} {\bibfnamefont {Jinshuang}\ \bibnamefont {Jin}}, \
  and\ \bibinfo {author} {\bibfnamefont {Mei}\ \bibnamefont {Huang}},\
  }\bibfield  {title} {\enquote {\bibinfo {title} {{Phase diagram and
  thermodynamics of the Polyakov linear sigma model with three quark
  flavors}},}\ }\href {\doibase 10.1088/0954-3899/37/3/035001} {\bibfield
  {journal} {\bibinfo  {journal} {J. Phys.}\ }\textbf {\bibinfo {volume}
  {G37}},\ \bibinfo {pages} {035001} (\bibinfo {year} {2010})},\ \Eprint
  {http://arxiv.org/abs/0906.1324} {arXiv:0906.1324 [hep-ph]} \BibitemShut
  {NoStop}%
\bibitem [{\citenamefont {Fisher}\ and\ \citenamefont
  {Barber}(1972)}]{Fisher:1972zza}%
  \BibitemOpen
  \bibfield  {author} {\bibinfo {author} {\bibfnamefont {Michael~E.}\
  \bibnamefont {Fisher}}\ and\ \bibinfo {author} {\bibfnamefont {Michael~N.}\
  \bibnamefont {Barber}},\ }\bibfield  {title} {\enquote {\bibinfo {title}
  {{Scaling Theory for Finite-Size Effects in the Critical Region}},}\ }\href
  {\doibase 10.1103/PhysRevLett.28.1516} {\bibfield  {journal} {\bibinfo
  {journal} {Phys. Rev. Lett.}\ }\textbf {\bibinfo {volume} {28}},\ \bibinfo
  {pages} {1516--1519} (\bibinfo {year} {1972})}\BibitemShut {NoStop}%
\bibitem [{\citenamefont {Abreu}\ \emph {et~al.}(2006)\citenamefont {Abreu},
  \citenamefont {Gomes},\ and\ \citenamefont {da~Silva}}]{Abreu:2006pt}%
  \BibitemOpen
  \bibfield  {author} {\bibinfo {author} {\bibfnamefont {L.~M.}\ \bibnamefont
  {Abreu}}, \bibinfo {author} {\bibfnamefont {M.}~\bibnamefont {Gomes}}, \ and\
  \bibinfo {author} {\bibfnamefont {A.~J.}\ \bibnamefont {da~Silva}},\
  }\bibfield  {title} {\enquote {\bibinfo {title} {{Finite-size effects on the
  phase structure of the Nambu-Jona-Lasinio model}},}\ }\href {\doibase
  10.1016/j.physletb.2006.10.015} {\bibfield  {journal} {\bibinfo  {journal}
  {Phys. Lett.}\ }\textbf {\bibinfo {volume} {B642}},\ \bibinfo {pages}
  {551--562} (\bibinfo {year} {2006})},\ \Eprint
  {http://arxiv.org/abs/hep-th/0610111} {arXiv:hep-th/0610111 [hep-th]}
  \BibitemShut {NoStop}%
\bibitem [{\citenamefont {Palhares}\ \emph {et~al.}(2011)\citenamefont
  {Palhares}, \citenamefont {Fraga},\ and\ \citenamefont
  {Kodama}}]{Palhares:2009tf}%
  \BibitemOpen
  \bibfield  {author} {\bibinfo {author} {\bibfnamefont {L.~F.}\ \bibnamefont
  {Palhares}}, \bibinfo {author} {\bibfnamefont {E.~S.}\ \bibnamefont {Fraga}},
  \ and\ \bibinfo {author} {\bibfnamefont {T.}~\bibnamefont {Kodama}},\
  }\bibfield  {title} {\enquote {\bibinfo {title} {{Chiral transition in a
  finite system and possible use of finite size scaling in relativistic heavy
  ion collisions}},}\ }\href {\doibase 10.1088/0954-3899/38/8/085101}
  {\bibfield  {journal} {\bibinfo  {journal} {J. Phys.}\ }\textbf {\bibinfo
  {volume} {G38}},\ \bibinfo {pages} {085101} (\bibinfo {year} {2011})},\
  \Eprint {http://arxiv.org/abs/0904.4830} {arXiv:0904.4830 [nucl-th]}
  \BibitemShut {NoStop}%
\bibitem [{\citenamefont {Fraga}\ \emph {et~al.}(2011)\citenamefont {Fraga},
  \citenamefont {Palhares},\ and\ \citenamefont {Sorensen}}]{Fraga:2011hi}%
  \BibitemOpen
  \bibfield  {author} {\bibinfo {author} {\bibfnamefont {Eduardo~S.}\
  \bibnamefont {Fraga}}, \bibinfo {author} {\bibfnamefont {Leticia~F.}\
  \bibnamefont {Palhares}}, \ and\ \bibinfo {author} {\bibfnamefont {Paul}\
  \bibnamefont {Sorensen}},\ }\bibfield  {title} {\enquote {\bibinfo {title}
  {{Finite-size scaling as a tool in the search for the QCD critical point in
  heavy ion data}},}\ }\href {\doibase 10.1103/PhysRevC.84.011903} {\bibfield
  {journal} {\bibinfo  {journal} {Phys. Rev.}\ }\textbf {\bibinfo {volume}
  {C84}},\ \bibinfo {pages} {011903} (\bibinfo {year} {2011})},\ \Eprint
  {http://arxiv.org/abs/1104.3755} {arXiv:1104.3755 [hep-ph]} \BibitemShut
  {NoStop}%
\bibitem [{\citenamefont {Bhattacharyya}\ \emph {et~al.}(2013)\citenamefont
  {Bhattacharyya}, \citenamefont {Deb}, \citenamefont {Ghosh}, \citenamefont
  {Ray},\ and\ \citenamefont {Sur}}]{Bhattacharyya:2012rp}%
  \BibitemOpen
  \bibfield  {author} {\bibinfo {author} {\bibfnamefont {Abhijit}\ \bibnamefont
  {Bhattacharyya}}, \bibinfo {author} {\bibfnamefont {Paramita}\ \bibnamefont
  {Deb}}, \bibinfo {author} {\bibfnamefont {Sanjay~K.}\ \bibnamefont {Ghosh}},
  \bibinfo {author} {\bibfnamefont {Rajarshi}\ \bibnamefont {Ray}}, \ and\
  \bibinfo {author} {\bibfnamefont {Subrata}\ \bibnamefont {Sur}},\ }\bibfield
  {title} {\enquote {\bibinfo {title} {{Thermodynamic Properties of Strongly
  Interacting Matter in Finite Volume using Polyakov-Nambu-Jona-Lasinio
  Model}},}\ }\href {\doibase 10.1103/PhysRevD.87.054009} {\bibfield  {journal}
  {\bibinfo  {journal} {Phys. Rev.}\ }\textbf {\bibinfo {volume} {D87}},\
  \bibinfo {pages} {054009} (\bibinfo {year} {2013})},\ \Eprint
  {http://arxiv.org/abs/1212.5893} {arXiv:1212.5893 [hep-ph]} \BibitemShut
  {NoStop}%
\bibitem [{\citenamefont {Bhattacharyya}\ \emph
  {et~al.}(2015{\natexlab{a}})\citenamefont {Bhattacharyya}, \citenamefont
  {Ray},\ and\ \citenamefont {Sur}}]{Bhattacharyya:2014uxa}%
  \BibitemOpen
  \bibfield  {author} {\bibinfo {author} {\bibfnamefont {Abhijit}\ \bibnamefont
  {Bhattacharyya}}, \bibinfo {author} {\bibfnamefont {Rajarshi}\ \bibnamefont
  {Ray}}, \ and\ \bibinfo {author} {\bibfnamefont {Subrata}\ \bibnamefont
  {Sur}},\ }\bibfield  {title} {\enquote {\bibinfo {title} {{Fluctuation of
  strongly interacting matter in the Polyakov–Nambu–Jona-Lasinio model in a
  finite volume}},}\ }\href {\doibase 10.1103/PhysRevD.91.051501} {\bibfield
  {journal} {\bibinfo  {journal} {Phys. Rev.}\ }\textbf {\bibinfo {volume}
  {D91}},\ \bibinfo {pages} {051501} (\bibinfo {year} {2015}{\natexlab{a}})},\
  \Eprint {http://arxiv.org/abs/1412.8316} {arXiv:1412.8316 [hep-ph]}
  \BibitemShut {NoStop}%
\bibitem [{\citenamefont {Magdy}\ \emph {et~al.}(2017)\citenamefont {Magdy},
  \citenamefont {Csanád},\ and\ \citenamefont {Lacey}}]{Magdy:2015eda}%
  \BibitemOpen
  \bibfield  {author} {\bibinfo {author} {\bibfnamefont {Niseem}\ \bibnamefont
  {Magdy}}, \bibinfo {author} {\bibfnamefont {M.}~\bibnamefont {Csanád}}, \
  and\ \bibinfo {author} {\bibfnamefont {Roy~A.}\ \bibnamefont {Lacey}},\
  }\bibfield  {title} {\enquote {\bibinfo {title} {{Influence of finite volume
  and magnetic field effects on the QCD phase diagram}},}\ }\href {\doibase
  10.1088/1361-6471/44/2/025101} {\bibfield  {journal} {\bibinfo  {journal} {J.
  Phys.}\ }\textbf {\bibinfo {volume} {G44}},\ \bibinfo {pages} {025101}
  (\bibinfo {year} {2017})},\ \Eprint {http://arxiv.org/abs/1510.04380}
  {arXiv:1510.04380 [nucl-th]} \BibitemShut {NoStop}%
\bibitem [{\citenamefont {Almasi}\ \emph {et~al.}(2017)\citenamefont {Almasi},
  \citenamefont {Pisarski},\ and\ \citenamefont {Skokov}}]{Almasi:2016zqf}%
  \BibitemOpen
  \bibfield  {author} {\bibinfo {author} {\bibfnamefont {Gabor}\ \bibnamefont
  {Almasi}}, \bibinfo {author} {\bibfnamefont {Robert}\ \bibnamefont
  {Pisarski}}, \ and\ \bibinfo {author} {\bibfnamefont {Vladimir}\ \bibnamefont
  {Skokov}},\ }\bibfield  {title} {\enquote {\bibinfo {title} {{Volume
  dependence of baryon number cumulants and their ratios}},}\ }\href {\doibase
  10.1103/PhysRevD.95.056015} {\bibfield  {journal} {\bibinfo  {journal} {Phys.
  Rev.}\ }\textbf {\bibinfo {volume} {D95}},\ \bibinfo {pages} {056015}
  (\bibinfo {year} {2017})},\ \Eprint {http://arxiv.org/abs/1612.04416}
  {arXiv:1612.04416 [hep-ph]} \BibitemShut {NoStop}%
\bibitem [{\citenamefont {Borsanyi}\ \emph {et~al.}(2014)\citenamefont
  {Borsanyi}, \citenamefont {Fodor}, \citenamefont {Hoelbling}, \citenamefont
  {Katz}, \citenamefont {Krieg},\ and\ \citenamefont
  {Szabo}}]{Borsanyi:2013bia}%
  \BibitemOpen
  \bibfield  {author} {\bibinfo {author} {\bibfnamefont {Szabocls}\
  \bibnamefont {Borsanyi}}, \bibinfo {author} {\bibfnamefont {Zoltan}\
  \bibnamefont {Fodor}}, \bibinfo {author} {\bibfnamefont {Christian}\
  \bibnamefont {Hoelbling}}, \bibinfo {author} {\bibfnamefont {Sandor~D.}\
  \bibnamefont {Katz}}, \bibinfo {author} {\bibfnamefont {Stefan}\ \bibnamefont
  {Krieg}}, \ and\ \bibinfo {author} {\bibfnamefont {Kalman~K.}\ \bibnamefont
  {Szabo}},\ }\bibfield  {title} {\enquote {\bibinfo {title} {{Full result for
  the QCD equation of state with 2+1 flavors}},}\ }\href {\doibase
  10.1016/j.physletb.2014.01.007} {\bibfield  {journal} {\bibinfo  {journal}
  {Phys. Lett.}\ }\textbf {\bibinfo {volume} {B730}},\ \bibinfo {pages}
  {99--104} (\bibinfo {year} {2014})},\ \Eprint
  {http://arxiv.org/abs/1309.5258} {arXiv:1309.5258 [hep-lat]} \BibitemShut
  {NoStop}%
\bibitem [{\citenamefont {Bazavov}\ \emph {et~al.}(2012)\citenamefont {Bazavov}
  \emph {et~al.}}]{Bazavov:2012jq}%
  \BibitemOpen
  \bibfield  {author} {\bibinfo {author} {\bibfnamefont {A.}~\bibnamefont
  {Bazavov}} \emph {et~al.} (\bibinfo {collaboration} {HotQCD}),\ }\bibfield
  {title} {\enquote {\bibinfo {title} {{Fluctuations and Correlations of net
  baryon number, electric charge, and strangeness: A comparison of lattice QCD
  results with the hadron resonance gas model}},}\ }\href {\doibase
  10.1103/PhysRevD.86.034509} {\bibfield  {journal} {\bibinfo  {journal} {Phys.
  Rev.}\ }\textbf {\bibinfo {volume} {D86}},\ \bibinfo {pages} {034509}
  (\bibinfo {year} {2012})},\ \Eprint {http://arxiv.org/abs/1203.0784}
  {arXiv:1203.0784 [hep-lat]} \BibitemShut {NoStop}%
\bibitem [{\citenamefont {Lenaghan}\ \emph {et~al.}(2000)\citenamefont
  {Lenaghan}, \citenamefont {Rischke},\ and\ \citenamefont
  {Schaffner-Bielich}}]{Lenaghan:2000ey}%
  \BibitemOpen
  \bibfield  {author} {\bibinfo {author} {\bibfnamefont {Jonathan~T.}\
  \bibnamefont {Lenaghan}}, \bibinfo {author} {\bibfnamefont {Dirk~H.}\
  \bibnamefont {Rischke}}, \ and\ \bibinfo {author} {\bibfnamefont {Jurgen}\
  \bibnamefont {Schaffner-Bielich}},\ }\bibfield  {title} {\enquote {\bibinfo
  {title} {{Chiral symmetry restoration at nonzero temperature in the SU(3)(r)
  x SU(3)(l) linear sigma model}},}\ }\href {\doibase
  10.1103/PhysRevD.62.085008} {\bibfield  {journal} {\bibinfo  {journal} {Phys.
  Rev.}\ }\textbf {\bibinfo {volume} {D62}},\ \bibinfo {pages} {085008}
  (\bibinfo {year} {2000})},\ \Eprint {http://arxiv.org/abs/nucl-th/0004006}
  {arXiv:nucl-th/0004006 [nucl-th]} \BibitemShut {NoStop}%
\bibitem [{\citenamefont {Schaefer}\ and\ \citenamefont
  {Wagner}(2009{\natexlab{b}})}]{Schaefer:2008hk}%
  \BibitemOpen
  \bibfield  {author} {\bibinfo {author} {\bibfnamefont {Bernd-Jochen}\
  \bibnamefont {Schaefer}}\ and\ \bibinfo {author} {\bibfnamefont {Mathias}\
  \bibnamefont {Wagner}},\ }\bibfield  {title} {\enquote {\bibinfo {title}
  {{The Three-flavor chiral phase structure in hot and dense QCD matter}},}\
  }\href {\doibase 10.1103/PhysRevD.79.014018} {\bibfield  {journal} {\bibinfo
  {journal} {Phys. Rev.}\ }\textbf {\bibinfo {volume} {D79}},\ \bibinfo {pages}
  {014018} (\bibinfo {year} {2009}{\natexlab{b}})},\ \Eprint
  {http://arxiv.org/abs/0808.1491} {arXiv:0808.1491 [hep-ph]} \BibitemShut
  {NoStop}%
\bibitem [{\citenamefont {Polyakov}(1978)}]{Polyakov:1978vu}%
  \BibitemOpen
  \bibfield  {author} {\bibinfo {author} {\bibfnamefont {Alexander~M.}\
  \bibnamefont {Polyakov}},\ }\bibfield  {title} {\enquote {\bibinfo {title}
  {{Thermal Properties of Gauge Fields and Quark Liberation}},}\ }\href
  {\doibase 10.1016/0370-2693(78)90737-2} {\bibfield  {journal} {\bibinfo
  {journal} {Phys. Lett.}\ }\textbf {\bibinfo {volume} {72B}},\ \bibinfo
  {pages} {477--480} (\bibinfo {year} {1978})}\BibitemShut {NoStop}%
\bibitem [{\citenamefont {Susskind}(1979)}]{Susskind:1979up}%
  \BibitemOpen
  \bibfield  {author} {\bibinfo {author} {\bibfnamefont {Leonard}\ \bibnamefont
  {Susskind}},\ }\bibfield  {title} {\enquote {\bibinfo {title} {{Lattice
  Models of Quark Confinement at High Temperature}},}\ }\href {\doibase
  10.1103/PhysRevD.20.2610} {\bibfield  {journal} {\bibinfo  {journal} {Phys.
  Rev.}\ }\textbf {\bibinfo {volume} {D20}},\ \bibinfo {pages} {2610--2618}
  (\bibinfo {year} {1979})}\BibitemShut {NoStop}%
\bibitem [{\citenamefont {Ratti}\ \emph {et~al.}(2006)\citenamefont {Ratti},
  \citenamefont {Thaler},\ and\ \citenamefont {Weise}}]{Ratti:2005jh}%
  \BibitemOpen
  \bibfield  {author} {\bibinfo {author} {\bibfnamefont {Claudia}\ \bibnamefont
  {Ratti}}, \bibinfo {author} {\bibfnamefont {Michael~A.}\ \bibnamefont
  {Thaler}}, \ and\ \bibinfo {author} {\bibfnamefont {Wolfram}\ \bibnamefont
  {Weise}},\ }\bibfield  {title} {\enquote {\bibinfo {title} {{Phases of QCD:
  Lattice thermodynamics and a field theoretical model}},}\ }\href {\doibase
  10.1103/PhysRevD.73.014019} {\bibfield  {journal} {\bibinfo  {journal} {Phys.
  Rev.}\ }\textbf {\bibinfo {volume} {D73}},\ \bibinfo {pages} {014019}
  (\bibinfo {year} {2006})},\ \Eprint {http://arxiv.org/abs/hep-ph/0506234}
  {arXiv:hep-ph/0506234 [hep-ph]} \BibitemShut {NoStop}%
\bibitem [{\citenamefont {Ghosh}\ \emph {et~al.}(2008)\citenamefont {Ghosh},
  \citenamefont {Mukherjee}, \citenamefont {Mustafa},\ and\ \citenamefont
  {Ray}}]{Ghosh:2007wy}%
  \BibitemOpen
  \bibfield  {author} {\bibinfo {author} {\bibfnamefont {Sanjay~K.}\
  \bibnamefont {Ghosh}}, \bibinfo {author} {\bibfnamefont {Tamal~K.}\
  \bibnamefont {Mukherjee}}, \bibinfo {author} {\bibfnamefont {Munshi~Golam}\
  \bibnamefont {Mustafa}}, \ and\ \bibinfo {author} {\bibfnamefont {Rajarshi}\
  \bibnamefont {Ray}},\ }\bibfield  {title} {\enquote {\bibinfo {title} {{PNJL
  model with a Van der Monde term}},}\ }\href {\doibase
  10.1103/PhysRevD.77.094024} {\bibfield  {journal} {\bibinfo  {journal} {Phys.
  Rev.}\ }\textbf {\bibinfo {volume} {D77}},\ \bibinfo {pages} {094024}
  (\bibinfo {year} {2008})},\ \Eprint {http://arxiv.org/abs/0710.2790}
  {arXiv:0710.2790 [hep-ph]} \BibitemShut {NoStop}%
\bibitem [{\citenamefont {Haas}\ \emph {et~al.}(2013)\citenamefont {Haas},
  \citenamefont {Stiele}, \citenamefont {Braun}, \citenamefont {Pawlowski},\
  and\ \citenamefont {Schaffner-Bielich}}]{Haas:2013qwp}%
  \BibitemOpen
  \bibfield  {author} {\bibinfo {author} {\bibfnamefont {Lisa~M.}\ \bibnamefont
  {Haas}}, \bibinfo {author} {\bibfnamefont {Rainer}\ \bibnamefont {Stiele}},
  \bibinfo {author} {\bibfnamefont {Jens}\ \bibnamefont {Braun}}, \bibinfo
  {author} {\bibfnamefont {Jan~M.}\ \bibnamefont {Pawlowski}}, \ and\ \bibinfo
  {author} {\bibfnamefont {Jürgen}\ \bibnamefont {Schaffner-Bielich}},\
  }\bibfield  {title} {\enquote {\bibinfo {title} {{Improved Polyakov-loop
  potential for effective models from functional calculations}},}\ }\href
  {\doibase 10.1103/PhysRevD.87.076004} {\bibfield  {journal} {\bibinfo
  {journal} {Phys. Rev.}\ }\textbf {\bibinfo {volume} {D87}},\ \bibinfo {pages}
  {076004} (\bibinfo {year} {2013})},\ \Eprint {http://arxiv.org/abs/1302.1993}
  {arXiv:1302.1993 [hep-ph]} \BibitemShut {NoStop}%
\bibitem [{\citenamefont {Schaefer}\ \emph {et~al.}(2009)\citenamefont
  {Schaefer}, \citenamefont {Wagner},\ and\ \citenamefont
  {Wambach}}]{Schaefer:2009st}%
  \BibitemOpen
  \bibfield  {author} {\bibinfo {author} {\bibfnamefont {Bernd-Jochen}\
  \bibnamefont {Schaefer}}, \bibinfo {author} {\bibfnamefont {Mathias}\
  \bibnamefont {Wagner}}, \ and\ \bibinfo {author} {\bibfnamefont {Jochen}\
  \bibnamefont {Wambach}},\ }\bibfield  {title} {\enquote {\bibinfo {title}
  {{QCD thermodynamics with effective models}},}\ }\bibfield  {booktitle}
  {\emph {\bibinfo {booktitle} {{Proceedings, 5th International Workshop on
  Critical point and onset of deconfinement (CPOD 2009): Upton, USA, June 8-12,
  2009}}},\ }\href@noop {} {\bibfield  {journal} {\bibinfo  {journal} {PoS}\
  }\textbf {\bibinfo {volume} {CPOD2009}},\ \bibinfo {pages} {017} (\bibinfo
  {year} {2009})},\ \Eprint {http://arxiv.org/abs/0909.0289} {arXiv:0909.0289
  [hep-ph]} \BibitemShut {NoStop}%
\bibitem [{\citenamefont {Tawfik}\ \emph {et~al.}(2014)\citenamefont {Tawfik},
  \citenamefont {Magdy},\ and\ \citenamefont {Diab}}]{Tawfik:2014uka}%
  \BibitemOpen
  \bibfield  {author} {\bibinfo {author} {\bibfnamefont {A.}~\bibnamefont
  {Tawfik}}, \bibinfo {author} {\bibfnamefont {N.}~\bibnamefont {Magdy}}, \
  and\ \bibinfo {author} {\bibfnamefont {A.}~\bibnamefont {Diab}},\ }\bibfield
  {title} {\enquote {\bibinfo {title} {{Polyakov linear SU(3) $\sigma$ model:
  Features of higher-order moments in a dense and thermal hadronic medium}},}\
  }\href {\doibase 10.1103/PhysRevC.89.055210} {\bibfield  {journal} {\bibinfo
  {journal} {Phys. Rev.}\ }\textbf {\bibinfo {volume} {C89}},\ \bibinfo {pages}
  {055210} (\bibinfo {year} {2014})},\ \Eprint {http://arxiv.org/abs/1405.0577}
  {arXiv:1405.0577 [hep-ph]} \BibitemShut {NoStop}%
\bibitem [{\citenamefont {Kovács}\ and\ \citenamefont
  {Wolf}(2015)}]{Kovacs:2015pha}%
  \BibitemOpen
  \bibfield  {author} {\bibinfo {author} {\bibfnamefont {Péter}\ \bibnamefont
  {Kovács}}\ and\ \bibinfo {author} {\bibfnamefont {György}\ \bibnamefont
  {Wolf}},\ }\bibfield  {title} {\enquote {\bibinfo {title} {{Chiral phase
  transition scenarios from the vector meson extended Polyakov quark meson
  model}},}\ }\bibfield  {booktitle} {\emph {\bibinfo {booktitle}
  {{Proceedings, International Meeting of Excited QCD 2015: Tatranska Lomnica,
  Slovakia, March 8-14, 2015}}},\ }\href {\doibase 10.5506/APhysPolBSupp.8.335}
  {\bibfield  {journal} {\bibinfo  {journal} {Acta Phys. Polon. Supp.}\
  }\textbf {\bibinfo {volume} {8}},\ \bibinfo {pages} {335} (\bibinfo {year}
  {2015})},\ \Eprint {http://arxiv.org/abs/1507.02064} {arXiv:1507.02064
  [hep-ph]} \BibitemShut {NoStop}%
\bibitem [{\citenamefont {Kovács}\ \emph {et~al.}(2016)\citenamefont
  {Kovács}, \citenamefont {Szép},\ and\ \citenamefont
  {Wolf}}]{Kovacs:2016juc}%
  \BibitemOpen
  \bibfield  {author} {\bibinfo {author} {\bibfnamefont {Peter}\ \bibnamefont
  {Kovács}}, \bibinfo {author} {\bibfnamefont {Zsolt}\ \bibnamefont {Szép}},
  \ and\ \bibinfo {author} {\bibfnamefont {György}\ \bibnamefont {Wolf}},\
  }\bibfield  {title} {\enquote {\bibinfo {title} {{Existence of the critical
  endpoint in the vector meson extended linear sigma model}},}\ }\href
  {\doibase 10.1103/PhysRevD.93.114014} {\bibfield  {journal} {\bibinfo
  {journal} {Phys. Rev.}\ }\textbf {\bibinfo {volume} {D93}},\ \bibinfo {pages}
  {114014} (\bibinfo {year} {2016})},\ \Eprint
  {http://arxiv.org/abs/1601.05291} {arXiv:1601.05291 [hep-ph]} \BibitemShut
  {NoStop}%
\bibitem [{\citenamefont {Fu}(2013)}]{Fu:2013ica}%
  \BibitemOpen
  \bibfield  {author} {\bibinfo {author} {\bibfnamefont {Wei-jie}\ \bibnamefont
  {Fu}},\ }\bibfield  {title} {\enquote {\bibinfo {title} {{Fluctuations and
  correlations of hot QCD matter in an external magnetic field}},}\ }\href
  {\doibase 10.1103/PhysRevD.88.014009} {\bibfield  {journal} {\bibinfo
  {journal} {Phys. Rev.}\ }\textbf {\bibinfo {volume} {D88}},\ \bibinfo {pages}
  {014009} (\bibinfo {year} {2013})},\ \Eprint {http://arxiv.org/abs/1306.5804}
  {arXiv:1306.5804 [hep-ph]} \BibitemShut {NoStop}%
\bibitem [{\citenamefont {Bhattacharyya}\ \emph
  {et~al.}(2015{\natexlab{b}})\citenamefont {Bhattacharyya}, \citenamefont
  {Ray}, \citenamefont {Samanta},\ and\ \citenamefont
  {Sur}}]{Bhattacharyya:2015zka}%
  \BibitemOpen
  \bibfield  {author} {\bibinfo {author} {\bibfnamefont {Abhijit}\ \bibnamefont
  {Bhattacharyya}}, \bibinfo {author} {\bibfnamefont {Rajarshi}\ \bibnamefont
  {Ray}}, \bibinfo {author} {\bibfnamefont {Subhasis}\ \bibnamefont {Samanta}},
  \ and\ \bibinfo {author} {\bibfnamefont {Subrata}\ \bibnamefont {Sur}},\
  }\bibfield  {title} {\enquote {\bibinfo {title} {{Thermodynamics and
  fluctuations of conserved charges in a hadron resonance gas model in a finite
  volume}},}\ }\href {\doibase 10.1103/PhysRevC.91.041901} {\bibfield
  {journal} {\bibinfo  {journal} {Phys. Rev.}\ }\textbf {\bibinfo {volume}
  {C91}},\ \bibinfo {pages} {041901} (\bibinfo {year} {2015}{\natexlab{b}})},\
  \Eprint {http://arxiv.org/abs/1502.00889} {arXiv:1502.00889 [hep-ph]}
  \BibitemShut {NoStop}%
\bibitem [{\citenamefont {Schaefer}\ \emph {et~al.}(2010)\citenamefont
  {Schaefer}, \citenamefont {Wagner},\ and\ \citenamefont
  {Wambach}}]{Schaefer:2009ui}%
  \BibitemOpen
  \bibfield  {author} {\bibinfo {author} {\bibfnamefont {Bernd-Jochen}\
  \bibnamefont {Schaefer}}, \bibinfo {author} {\bibfnamefont {Mathias}\
  \bibnamefont {Wagner}}, \ and\ \bibinfo {author} {\bibfnamefont {Jochen}\
  \bibnamefont {Wambach}},\ }\bibfield  {title} {\enquote {\bibinfo {title}
  {{Thermodynamics of (2+1)-flavor QCD: Confronting Models with Lattice
  Studies}},}\ }\href {\doibase 10.1103/PhysRevD.81.074013} {\bibfield
  {journal} {\bibinfo  {journal} {Phys. Rev.}\ }\textbf {\bibinfo {volume}
  {D81}},\ \bibinfo {pages} {074013} (\bibinfo {year} {2010})},\ \Eprint
  {http://arxiv.org/abs/0910.5628} {arXiv:0910.5628 [hep-ph]} \BibitemShut
  {NoStop}%
\end{thebibliography}%
\end{document}